\documentclass[12pt]{iopart}
 \expandafter\let\csname equation*\endcsname\relax
  \expandafter\let\csname endequation*\endcsname\relax
\usepackage{graphicx,bbm,color,amsmath}
\usepackage{dsfont, eufrak}
\let\mathfrak\undefined
\usepackage{amsfonts}
\usepackage{dsfont}
\usepackage{calligra}
\usepackage{calrsfs}
\usepackage{float}
\usepackage[mathscr]{euscript}
\usepackage[utf8]{inputenc} 

\newcommand{\be}{\begin{equation}}
\newcommand{\ee}{\end{equation}}

\newcommand{\ii}{ {\rm i} }

\newcommand{\ZZ}{\mathbb{Z}}

\begin{document}

\title[Corner transfer matrices for 2D strongly coupled...]{Corner transfer matrices for 2D strongly coupled many-body Floquet systems} 

\author{Ivan Kukuljan and Toma\v{z} Prosen}

\address{Department of Physics, Faculty of Mathematics and Physics, University of Ljubljana, Jadranska ulica 19, SI-1000 Ljubljana, Slovenia}

\begin{abstract}
We develop, based on Baxter's corner transfer matrices, a renormalizable numerically exact method for computation of the level density of the quasi-energy spectra of two-dimensional (2D) locally interacting many-body Floquet systems. We demonstrate its functionality exemplified by the kicked 2D quantum Ising model. Using the method, we are able to treat the system of arbitrarily large finite size (for example $10000\times10000$ lattice). We clearly demonstrate that the density of Floquet quasi-energy spectrum tends to a flat function in the thermodynamic limit for generic values of model parameters. However, contrary to the prediction of random matrices of the circular orthogonal ensemble, the decay rates of the Fourier coefficients of the Floquet level density exhibit rich and non-trivial dependence on the system's parameters. Remarkably, we find that the method is renormalizable and gives thermodynamically convergent results only in certain regions of the parameter space where the corner transfer matrices have effectively a finite rank for any system size.
In the complementary regions, the corner transfer matrices effectively become of full rank and the method becomes non-renormalizable. This may indicate an interesting phase transition from an area- to volume- law of entanglement in the thermodynamic state of a Floquet system.
\end{abstract}
\vspace{10 mm}
\maketitle

\section{Introduction}

Understanding dynamical features of strongly correlated quantum many-body systems is one of the central challenges of contemporary theoretical and experimental condensed matter physics. While interacting quantum systems in one spatial dimension seem to be efficiently simulable due to their moderate entanglement content encapsulated in the so-called {\em area law} \cite{Eisert2010}, some of the main unsolved theoretical challenges of condensed matter physics, such as e.g. high $T_c$ superconductivity, could be connected to intractability of strongly correlated quantum systems in two spatial dimensions. In order to elucidate the principal nontrivial quantum dynamical phenomena, it is often beneficial to define and study the simplest possible mathematical models which display certain effects. In this sense, dynamical and non-equilibrium phenomena could most clearly be illustrated and understood in the so-called Floquet systems, i.e. time-dependent (driven) Hamiltonian quantum systems (see e.g. \cite{Haake}). Such systems have no a-priori conservation laws in a generic (nonintegrable) case and dynamics can be reduced to a discrete quantum dynamical system generated by the so-called unitary Floquet propagator over one period of the driving.

Even though Floquet systems have been proposed by one of us as paradigmatic models of quantum many-body dynamics and possible candidates exhibiting nontrivial ergodicity--nonergodicity transitions already almost two decades ago \cite{ProsenPRL1998,ProsenJPA1998,ProsenPRE1999}, see also Ref.~\cite{ProsenPRE2002} for the connection to Loschmidt echo, they have only very recently generated a renewed interest in the general context of nonequilibrium dynamics \cite{Polkovnikov,Rigol} and potential experimental relevance in cold atomic setups. Moreover, remarkable resistivity of high-frequency Floquet systems against heating \cite{Demler} can at least partially be understood in light of the recent results and rigorous theorems on asymptotic convergence of the Magnus expansion \cite{Kuwahara2015,Mori2015,Abanin2015}. Such an emergent Floquet non-ergodicity can also be interpreted as {\em many-body-localization} in the absence of disorder.

Recently, numerical experiments with Floquet map of a quantum Ising model on a finite 2D square spin lattice periodically kicked with a transverse magnetic field suggested an intriguing behavior of the simplest dynamical quantity one can imagine: the Floquet level density (the quasienergy spectrum) \cite{PinedaProsen2014}. Namely, it has been suggested that, depending on the model parameters, the Fourier modes of the ($2\pi-$periodic) Floquet level density do not converge to zero, or converge to zero significantly more slowly than would be expected from random matrix theory which may naively be expected to describe such strongly non-integrable quantum system. This means that deviations of the Floquet level density from a uniform one can be much larger than would be expected based on the corresponding maximum entropy random matrix ensemble 
(in this case, circular orthogonal ensemble (COE) \cite{MehtaRM}). However, the results of Ref.~\cite{PinedaProsen2014} have been based on brute force numerical simulations of small spin lattices (up to $5\times 5$) and a critical question has remained whether they have any relevance for the behavior of the system in the thermodynamic limit.

In this work we address precisely this question. To resolve it, we modify and implement the {\em corner transfer matrices} (CTM) method for the computation of Floquet level density of periodically driven quantum spin systems on 2D square-based lattices. CTM was developed by Rodney J. Baxter in the end of 1960s and has been mostly (but not solely) used in classical statistical physics. It can be understood as a version of numerical renormalization group scheme. Our implementation enables numerically exact computation of the Fourier modes of the Floquet level density of the system with arbitrarily large finite size (we went up to $10000\times10000$ lattices) with no simplifications of the system required. The only source of error is the truncation of insignificant (tiny) eigenvalues of corner transfer matrices.

The method is shown to work efficiently in parts of the parameter space where the corner transfer matrices have an effectively finite rank. This can be heuristically interpreted as a kind of area-law property of a certain auxiliary many-body state of a two-dimensional lattice which enables exact computation within a polynomial time. 
As a result, we obtain accurate exponential decay of low Fourier modes in the thermodynamic limit indicating asymptotically uniform level density, with the rates which depend on the precise values of model parameters and cannot be described in terms of universal concepts such as random matrix theories.
For small lattice sizes, our results confirm the picture based on finite system analysis \cite{PinedaProsen2014}. In the complementary part of parameter space, the corner transfer matrices have a full rank and hence the method cannot access the thermodynamic limit.
This may also suggest an interesting new type of quantum phase transition from area- to volume-like complexity (entanglement) scaling in computation of short time dynamics (Fourier modes of Floquet level density being just a generic example) of quantum 2D lattice systems.

The paper is organized as follows. In section \ref{sec:2DKI} we define the 2D kicked quantum Ising model and outline the problem of computation of level density of the Floquet quasienergy spectrum.
In section \ref{sec:Method} we describe the CTM evaluation of the tensor network contraction which corresponds to evaluating the low Fourier modes of the Floquet level density.
This is followed by presentation and interpretation of the numerical results, demonstration of the robustness of the method (section \ref{sec:Results}) and conclusions (section \ref{sec:Conc}).

\section{Two dimensional kicked quantum Ising model and Floquet level density}\label{sec:2DKI}

\begin{figure}[H]
\centering
\hspace{1.2cm}\includegraphics[width=0.8\linewidth]{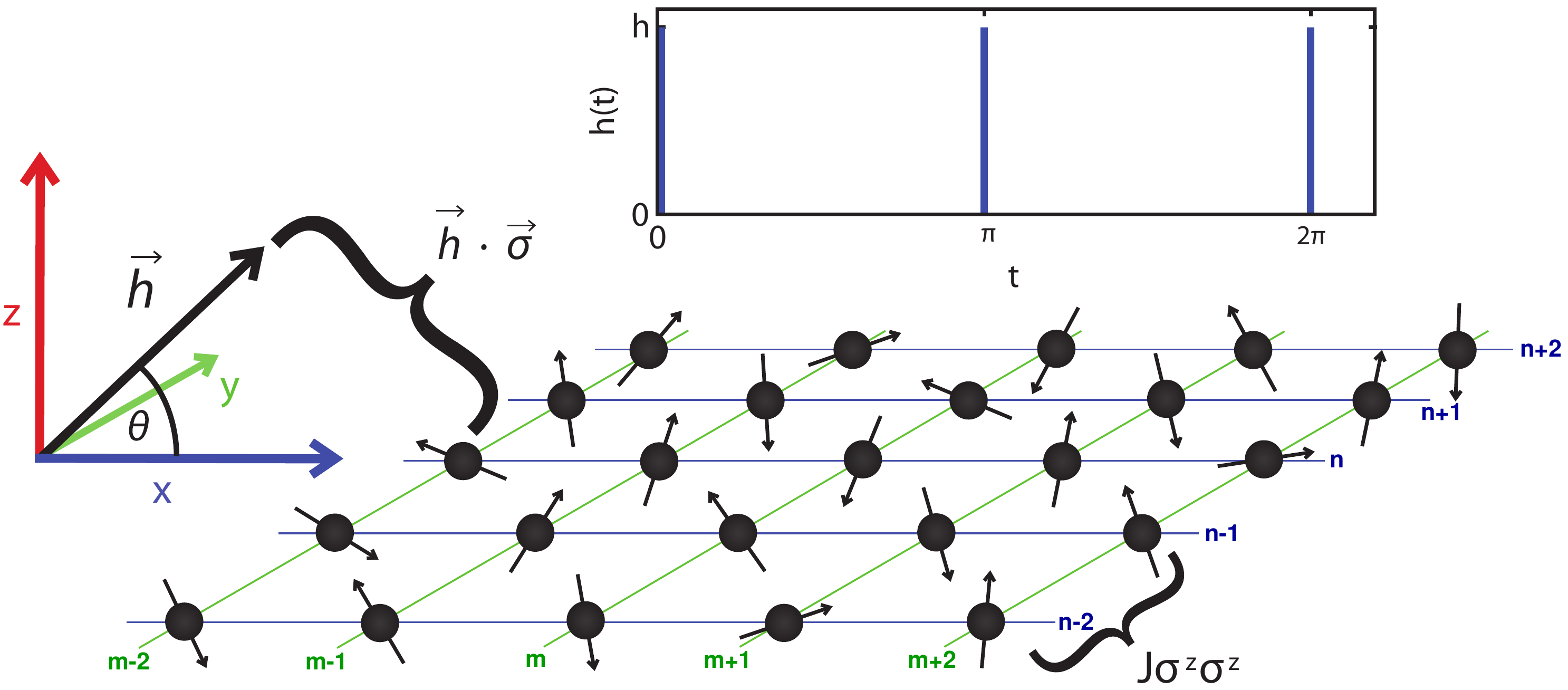}
\caption[Kicked 2D quantum Ising model.]{A symbolic illustration of the kicked 2D quantum Ising model. The spins are located in a square-based lattice $\Lambda\in\ZZ^2$ denoted with green and blue lines. The energy of the pair interaction of nearest-neighbor spins is given by the Ising term $J\sigma^{z}\sigma^{z}$ with $J$ the coupling constant. The system is periodically driven by a homogeneous external magnetic field $\vec{h} =\left(h\cos\left(\theta\right),0,h\sin\left(\theta\right)\right)$. It acts on the system with periodical $\delta$-kicks via the Zeeman coupling term $\vec{h}\cdot\vec{\sigma}$. The full Hamiltonian of the system is given by equation \eqref{eq:H}.}
\label{fig:2DKickedIsing}
\end{figure}

In a recent work \cite{PinedaProsen2014}, Pineda {\em et al.} proposed and studied the probably simplest non-trivial quantum dynamical many-body model in two dimensions, the 2D periodically kicked quantum Ising spin $1/2$ model (2D KI). Its Hamiltonian is given by:
\begin{equation}
H\left(t\right)=H_\text{Ising}+H_\text{kick}\sum_{j\in\mathbb{Z}}\delta\left(t-j\tau\right),\label{eq:H}
\end{equation}
where $H_\text{Ising}$ is the Hamiltonian of the autonomous 2D quantum Ising model with nearest neighbor coupling $J$:
\begin{equation}
H_\text{Ising}=J\sum_{m,n\in\Lambda}\left(\sigma_{m,n}^{z}\sigma_{m+1,n}^{z}+\sigma_{m,n}^{z}\sigma_{m,n+1}^{z}\right),\label{eq:H1}
\end{equation}
and $H_\text{kick}$ is the Zeeman term determining the interaction with the external spatially homogeneous magnetic field $\vec{h}$ with magnitude $h$ and angle $\theta$ with respect to transversal $x$-axis:
\begin{equation}
H_\text{kick}=\sum_{m,n\in\Lambda}\vec{h}\cdot\vec{\sigma}_{m,n}=h\sum_{m,n\in\Lambda}\left(\sigma_{m,n}^{x}\cos\theta+\sigma_{m,n}^{z}\sin\theta\right).\label{eq:H0}
\end{equation}
Here, $\vec{\sigma}$ are the standard Pauli matrices. The spins are ordered in a square-based 2D lattice $\Lambda\subseteq\ZZ^2$ of the desired shape and boundary conditions. In \cite{PinedaProsen2014}, rectangle-shaped lattice with periodic boundary conditions was considered. Because of the characteristics of our numerical method, we will consider a diamond-shaped lattice with open boundary conditions in this paper (like in figure \ref{fig:TensCirc1}). The external magnetic field acts on the system with periodical $\delta$-kicks with period $\tau$ (we set $\tau=1$ by choice of units). They can be easily generalized to arbitrary periodic functions, but $\delta$-pulses are sufficient to unveil the principles of periodically driven dynamics. Furthermore, $\vec{h}$ can be without loss of generality chosen to lie in the $x$-$z$ plane (second equality in equation \eqref{eq:H0}): $\vec{h}=\left(h_x,0,h_z\right)=\left(h\cos\left(\theta\right),0,h\sin\left(\theta\right)\right)$. Then $H$ is a real matrix in the eigenbasis of $\sigma_{m,n}^{z}$. In \cite{PinedaProsen2014} and in this paper, the transversal field ($\theta=0$) is mostly used. The 2D KI model is symbolically illustrated in figure \ref{fig:2DKickedIsing}.

\subsection{Level density of the quasienergy spectrum}


The most surprising results of Pineda {\em et al.} \cite{PinedaProsen2014} have been obtained by observing the spectrum of the Floquet operator of the kicked 2D KI model. They have indicated that in the thermodynamic limit Floquet level density does not converge to a constant function, or converges significantly more slowly than would be expected from the corresponding maximum entropy random matrix ensemble.

The \textit{Floquet map (operator)} $\mathcal{U}$ \cite{Haake} of a system with periodically time dependent Hamiltonian $H\left(t+n\tau\right)=H\left(t\right)$ is defined as the time evolution propagator over one period:
\begin{equation}
\mathcal{U}:=U\left(\tau\right)=T\left\{ \exp\left[-\frac{\ii}{\hbar}\int_0^\tau H\left(t\right)dt\right]\right\},
\end{equation}
where $T$ is the time ordering operator. We set $\hbar=1$. In the case of 2D KI model the Floquet map reads:
\begin{equation}
\mathcal{U}_{\text{KI}}=\mathcal{U}_{\text{Ising}}\mathcal{U}_{\text{kick}}=\exp\left(-\ii H_\text{Ising}\right)\exp\left(-\ii H_\text{kick}\right). \label{eq:FloqKI}
\end{equation} 
The \textit{spectrum} of the Floquet map \cite{Haake,PinedaProsen2014}, also called the \textit{quasienergy spectrum} of the system, $ \mathcal{S}=\left\{ \phi_{n};n=1,\ldots,\mathcal{N}:=2^{\left|\Lambda\right|}\right\}$, with $\left|\Lambda\right|$ the number of spins in the system, is defined by the unitary eigenvalue problem:
 \begin{equation}
 \mathcal{U}\left|\psi_{n}\right\rangle =e^{-\ii\phi_{n}}\left|\psi_{n}\right\rangle. \label{eq:quas}
 \end{equation}
Statistical properties of quasienergy spectra of single particle systems are among the most often studied objects in Quantum Chaos Theory (e.g. \cite{Haake}). However, the properties of the Floquet operators of many-body systems are only beginning to be intensively studied and they are still not very well understood (see e.g. \cite{PinedaProsen2014,Polkovnikov,Rigol,Demler,Kuwahara2015,Mori2015,Abanin2015}). We will be focusing on the simplest one among them, the \textit{one-point function}, or the Floquet spectral (or level) density:
\begin{equation}
\rho\left(\phi\right):=\frac{1}{\mathcal{N}}\sum_{n=1}^{\mathcal{N}}\delta\left(\phi-\phi_{n}\right). \label{eq:one-point}
\end{equation}
The level density $\rho\left(\phi\right)$ is defined on compact domain $\left[0,2\pi\right)$ and can be therefore expanded in a Fourier series:
\begin{equation}
\rho\left(\phi\right)=\frac{1}{2\pi}\left(1+2\sum_{k=1}^{\infty}\rho_{k}\cos\left(k\phi\right)\right),
\end{equation}
with Fourier coefficients:
\begin{equation}
\rho_{k}:=\int_{0}^{2\pi}\rho\left(\phi\right)e^{ik\phi}d\phi=\frac{1}{\mathcal{N}}\text{tr}\mathcal{U}^{k}. \label{eq:FourCoef}
\end{equation}
The last equality is obtained using definitions \eqref{eq:quas} and \eqref{eq:one-point}. It is an elegant expression that links the Fourier coefficients of the quasienergy spectrum to the powers of the Floquet operator which correspond to the propagation of the system over $k$ kicks (drive periods).  Our goal is to evaluate this expression for $\mathcal{U}_{\text{KI}}$ in the thermodynamic limit $\Lambda\rightarrow\ZZ^2$ (i.e. $\left|\Lambda\right|\rightarrow\infty$) and study its thermodynamic scaling. 

\subsection{Theoretical prediction for the thermodynamic limit}
Because 2D KI model is strongly nonintegrable and presumably highly chaotic, according to the Quantum Chaos Conjecture, the behavior of the Fourier coefficients $\rho_{k}$ in the thermodynamic limit may be predicted using {\em random matrix theory} (RMT) \cite{Haake,MehtaRM}. Due to unitarity and time-reversal symmetry of $\mathcal{U}_{\text{KI}}$, the corresponding random matrix ensemble is Dyson's circular orthogonal ensemble (COE) \cite{MehtaRM,Haake}. The expected value of $\rho_{k}$ in the thermodynamic limit can be found by evaluation of expression \eqref{eq:FourCoef} using $\text{tr}\mathcal{U}^{k}=\sum_{j=1}^{\mathcal{N}}e^{-i\phi_{j}k}$, $\mathcal{N}=2^{\left|\Lambda\right|}=2^{N^2}$, and the COE one-level correlation function (level density) \cite{MehtaRM,Haake}. One obtains a trivial result:
\begin{equation}
\left\langle \rho_{k}\right\rangle_{\text{COE}\left(\mathcal{N}\right)}=\delta_{k,0}.
\end{equation}
Therefore we expect all Fourier modes $\rho_{k}$ to decay in the thermodynamic limit and the level density of the quasienergy spectrum of 2D KI model to converge to a flat function. One may thus predict that the nontrivial phase diagram $\rho_{k}\left(h,J\right)$ found in \cite{PinedaProsen2014} should be a result of strong finite-size effects. Assuming that $\mathcal{U}$ can be treated as a {\em typical} member of $\mathrm{COE}({\cal N})$,
the asymptotic dependence $\rho_{k}$ on the system size $N$ for large $N$ can be estimated in terms of typical expected fluctuations (standard, root-mean-square deviation) within the ensemble as
\begin{equation}
\rho_{k} \sim\sqrt{\left\langle \rho_{k}^{2}\right\rangle _{\text{COE}\left(\mathcal{N}\right)}}.
\end{equation}
Using the COE two-level correlation function \cite{MehtaRM,Haake}, we obtain:
\begin{equation}
\rho_{k} \propto\frac{1}{\mathcal{N}}=2^{-\left|\Lambda\right|}. \label{eq:RMTprediction}
\end{equation}
Thus, according to a universal prediction of RMT, we should expect the Fourier coefficients $\rho_{k}$ to decay exponentially with the number of spins $\left|\Lambda\right|$ with the exponent $\log 2$. RMT also predicts the decay rate to be independent of the system's parameters $h$, $J$ and $\theta$. We shall demonstrate later that a generic dynamical many-body model, such as 2D KI, displays a different, non-universal behavior of the scaling of $\rho_k$ on $|\Lambda|$.

\section{Method}
\label{sec:Method}
The computational cost of evaluating expression \eqref{eq:FourCoef} in the straight-forward manner increases exponentially with the number of spins $\left|\Lambda\right|$ as the dimension of the Hilbert space in which the trace is to be taken is $\mathcal{N}=2^{\left|\Lambda\right|}$. So, a more sophisticated approach is required to reach large numbers of spins $\left|\Lambda\right|$.

\begin{figure}[H]
\centering
\hspace{1.2cm}\includegraphics[width=0.8\linewidth]{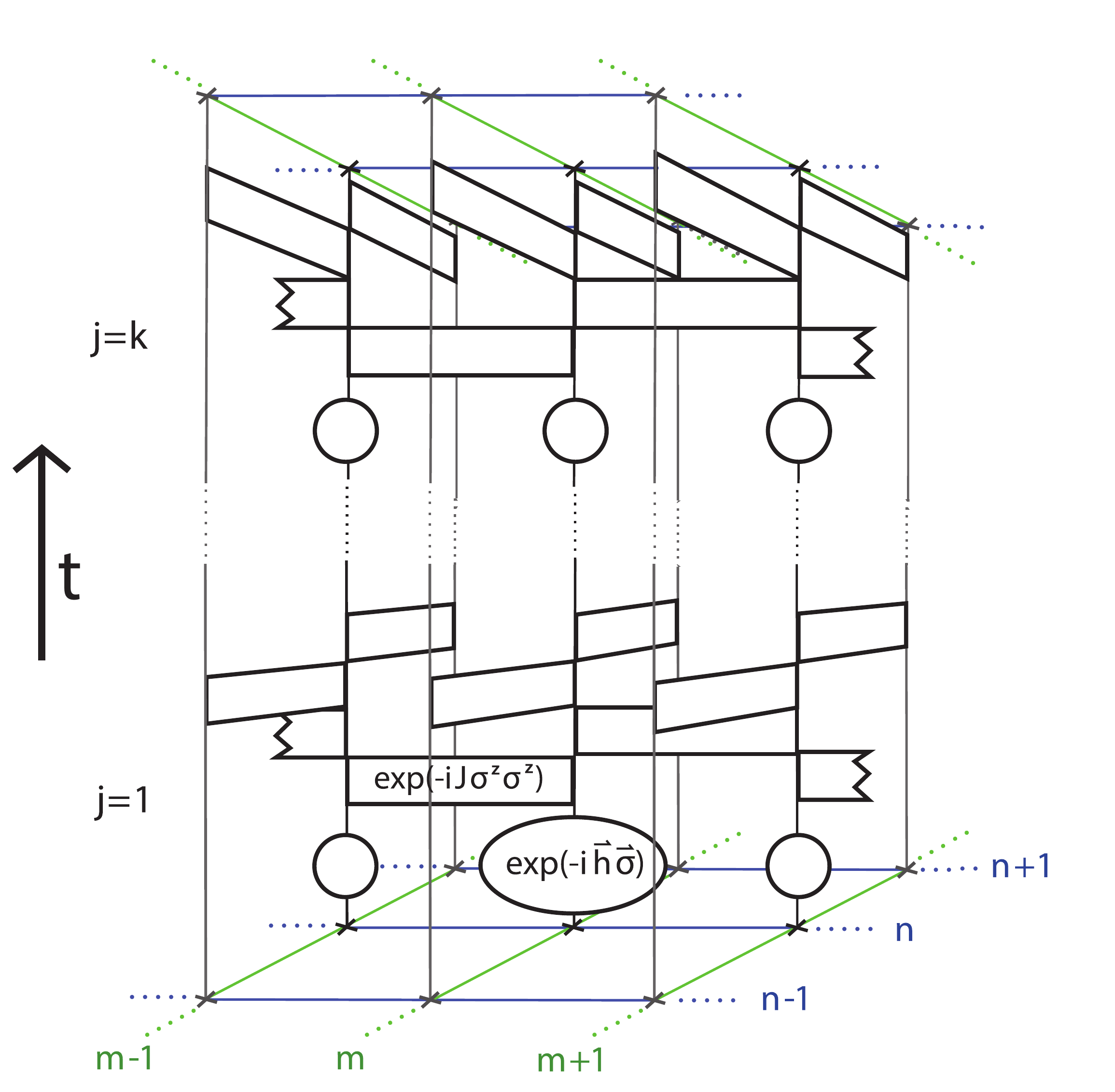}
\caption{The expression \eqref{eq:FourCoef} for the $k-$th Fourier mode of Floquet level density can be represented as the contraction of the 3D tensor network illustrated above. The physical spins are located on a square-based lattice $\Lambda\in\ZZ^2$ denoted by green and blue horizontal lines. Here, only a slice $\left\{m-1,m,m+1\right\}\times\left\{n-1,n,n+1\right\}$ of the lattice is presented. The vertical gray lines denote quantum-circuit wires corresponding to time-evolution of physical spins. They are crossed with circles and rectangles that denote, correspondingly, the one-spin and two-spin operators from $\mathcal{U}_{\text{kick}}$ and  $\mathcal{U}_{\text{Ising}}$ acting on them.  Floquet map $\mathcal{U}_{\text{KI}}$ acts on the system $k$ times, so we have $k$ repeated levels of operators. Contraction (taking the trace in expression \eqref{eq:FourCoef}) is usually denoted by joining the beginning and the end of each spin (closing the loops). For transparency, this is denoted by small crosses at the beginnings and the ends of the spins here. The spins have to be contracted simultaneously which corresponds to evaluating the trace over $\mathcal{N}=2^{\left|\Lambda\right|}$ dimensional Hilbert space.}
\label{fig:3DTensNet}
\end{figure}

Because the terms $\vec{h}\cdot\vec{\sigma}$ in $H_\text{kick}$ (eq. \eqref{eq:H0}) commute with each other when they act on different spins and the terms $\sigma^z\otimes\sigma^z$ in $H_\text{Ising}$ (eq. \eqref{eq:H1}) commute with each other, the Floquet map of 2D KI (eq. \eqref{eq:FloqKI}) factorizes in terms of one-spin and two-spin unitaries:
\begin{equation}
\mathcal{U}_{\text{KI}}=\prod_{m,n\in\Lambda}\exp\left(-\ii \sigma_{m,n}^{z}\sigma_{m+1,n}^{z}\right)\exp\left(-\ii \sigma_{m,n}^{z}\sigma_{m,n+1}^{z}\right)\exp\left(-\ii \vec{h}\cdot\vec{\sigma}_{m,n}\right) \label{eq:FactorizedFloquet}
\end{equation}
Using equation \eqref{eq:FactorizedFloquet}, the scalar expression \eqref{eq:FourCoef} can be represented as the contraction of the 3D tensor network shown in figure \ref{fig:3DTensNet}.

\begin{figure}[H]
	\centering
	\includegraphics[width=0.45\linewidth]{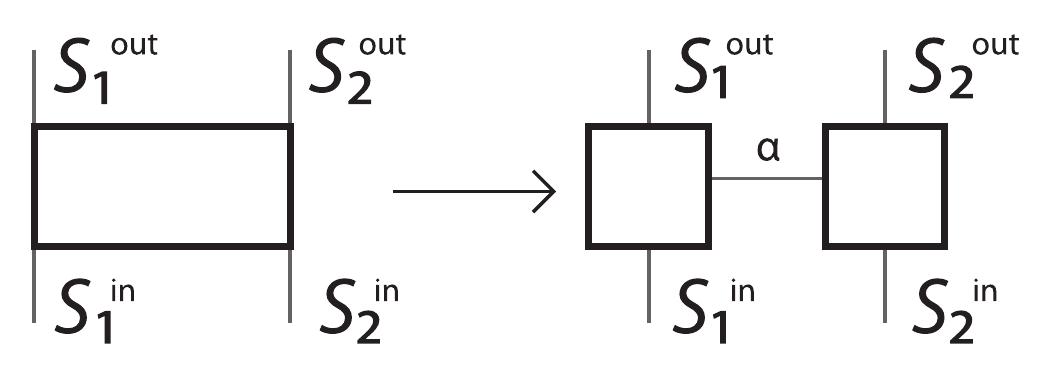}
	\caption[Decomposition of two-spin operators into products of one-spin operators.]{Two-spin operators from figure \ref{fig:3DTensNet} can be decomposed into products of two one-spin operators with an ancillary spin as the summation variable. Details are given in the main text.}
	\label{fig:TwoOneDec}
\end{figure}

The two-spin operators from figure \ref{fig:3DTensNet} can be decomposed as products of one-spin operators (figure \ref{fig:TwoOneDec}). Initially, the two-spin operators $O=\exp\left(-iJ\sigma^z\otimes\sigma^z\right)$ are $4\times4$ matrices with elements labeled as $O\left(s_1^\text{out}s_2^\text{out}|s_1^\text{in}s_2^\text{in}\right)$. So, the rows are labeled with the outgoing values $s_1^\text{out},s_2^\text{out}\in\left\{0,1\right\}$ and the columns with the incoming values $s_1^\text{in},s_2^\text{in}\in\left\{0,1\right\}$ of the two spins (in the computational basis, i.e., eigenbasis of $\sigma^z_{m,n}$). We rearrange the elements in the matrix to be ordered as
\begin{equation}
\tilde{O}\left(s_1^\text{out}s_1^\text{in}|s_2^\text{out}s_2^\text{in}\right)=O\left(s_1^\text{out}s_2^\text{out}|s_1^\text{in}s_2^\text{in}\right)
\end{equation}
so that the rows are labeled with the values of the first spin and the columns with the incoming and the outgoing value of the second spin. Then we compute the eigenvalue decomposition of a complex symmetric matrix 
\begin{equation}
\tilde{O}=VDV^{\top}.
\end{equation}
It turns out that there are only two non-zero eigenvalues so that we can keep only two rows and columns of $D$ and only two columns of $V$. We thus define a $4\times 2$ matrix:
\begin{equation}
\tilde{o}:=V\sqrt{D}.
\end{equation}
We can regard the two columns as belonging to the two different values of an ancillary spin $\alpha$ so that the elements are labeled $\tilde{o}\left(s^{\text{out}}s^{\text{in}}|\alpha\right)$. Then, the the two-spin operator $\tilde{O}$ can be written as a product of two equal one-spin operators $\tilde{o}$, $\tilde{O}=\tilde{o}\tilde{o}^{\top}$, where the summation is over the two possible values of an ancillary spin (figure \ref{fig:TwoOneDec}). For the further use we rearrange the elements of $\tilde{o}$ yielding a pair of $2\times2$ matrices $o^{\alpha}$ for $\alpha=0,1$:
\begin{equation}
o^{\alpha}\left(s^{\text{out}}|s^{\text{in}}\right)=\tilde{o}\left(s^{\text{out}}s^{\text{in}}|\alpha\right).
\end{equation}

\begin{figure}[H]
\centering
\hspace{1.8cm}\includegraphics[width=0.8\linewidth]{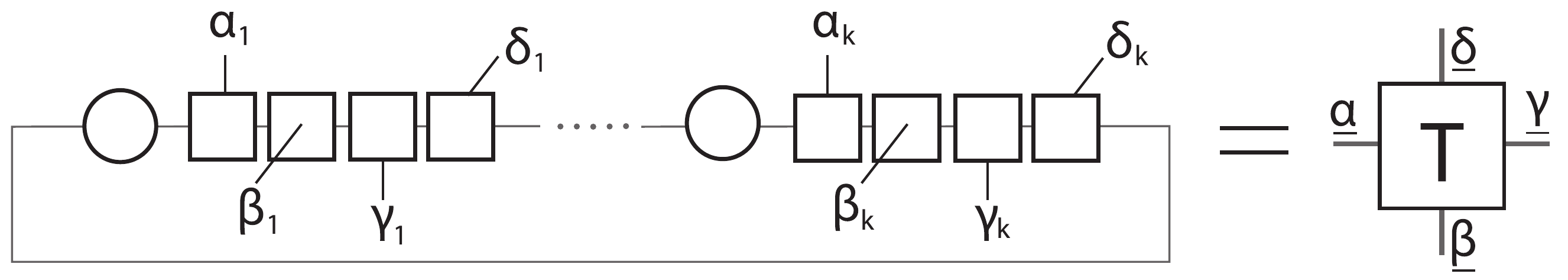}
\caption[Contraction of physical spins.]{The sketch is rotated by quarter of a turn to occupy less space on the page. After the decomposition of the two-spin operators into the products of one-spin operators (figure \ref{fig:TwoOneDec}), each physical spin from figure \ref{fig:3DTensNet} is connected to other physical spins only by ancillary spins (denoted by Greek letters). The physical spins can now be contracted individually. All the ancillary spins between the same pair of physical spins can be viewed as a single spin with $2^k$ possible values (denoted eith thicker lines). All the operators on the physical spin can be treated as a single operator of the connecting $2^k$-valued ancillary spins. The 3D tensor network has thus been reduced to a 2D tensor network.}
\label{fig:SpinContract}
\end{figure}

Decomposing two-spin operators into one-spin operators allows to contract (evaluate the trace over) each spin in the lattice in figure \ref{fig:3DTensNet} individually instead of having to compute the trace over the basis of the entire Hilbert space (figure \ref{fig:SpinContract}). In this way we reduce the 3D tensor network to a 2D tensor network - after having contracted the physical spins, we can view all the ancillary spins $\alpha_j$, $j=1,\ldots,k$, connecting the same pair of lattice sites (physical spins) as a single composite ancillary spin $\underline{\alpha}:=\left(\alpha_1\alpha_2\cdots\alpha_k\right)$ with $2^k$ possible values. In the figures \ref{fig:TensCirc1} and \ref{fig:TensCirc2} they are denoted by thicker gray lines. All the operators on each physical spin are joint into a single composite rank-4 tensor $T$ operating on $2^k$-valued ancillary spins. Expressed symbolically:
\begin{equation}
T^{\left(k\right)}\left(\underline{\alpha},\underline{\beta},\underline{\gamma},\underline{\delta}\right):=\frac{1}{2}\text{tr}\left(\prod_{j=1}^{k}o^{\alpha_{j}}o^{\beta_{j}}o^{\gamma_{j}}o^{\delta_{j}}\,a\right), \label{eq:tensorT}
\end{equation}
where $a:=\exp\left(-\ii \vec{h}\cdot\vec{\sigma}\right)$, and operator order in the symbolic product is here and below understood as $\prod_{j=1}^k B_j \equiv B_k B_{k-1} \cdots B_1$. The factor $\frac{1}{2}$ is to account for normalization: there are $\left|\Lambda\right|$ tensors or contracted spins (or tensors) in the system so in this way we incorporate the overall factor $\frac{1}{\mathcal{N}}=2^{-\left|\Lambda\right|}$ from equation \eqref{eq:FourCoef} already in the definition of the tensor network and do not have to worry about it later.
\subsection{Corner transfer matrices}
Contraction of the obtained 2D tensor network is done by summation over all the possible values of the ancillary spins. This is computationally equally expensive as contraction of the original 3D circuit. However, the 2D network allows for application of corner transfer matrices \cite{BaxterESM,BaxterCTM1981} which in turn enable us to contract lattices of arbitrary number of spins. Here we will present a reformulation of CTM method which is better suited to tensor networks than the original version, which is made for classical spin systems with an interaction around-a-face.

\begin{figure}[H]
\centering
\includegraphics[width=0.71\linewidth]{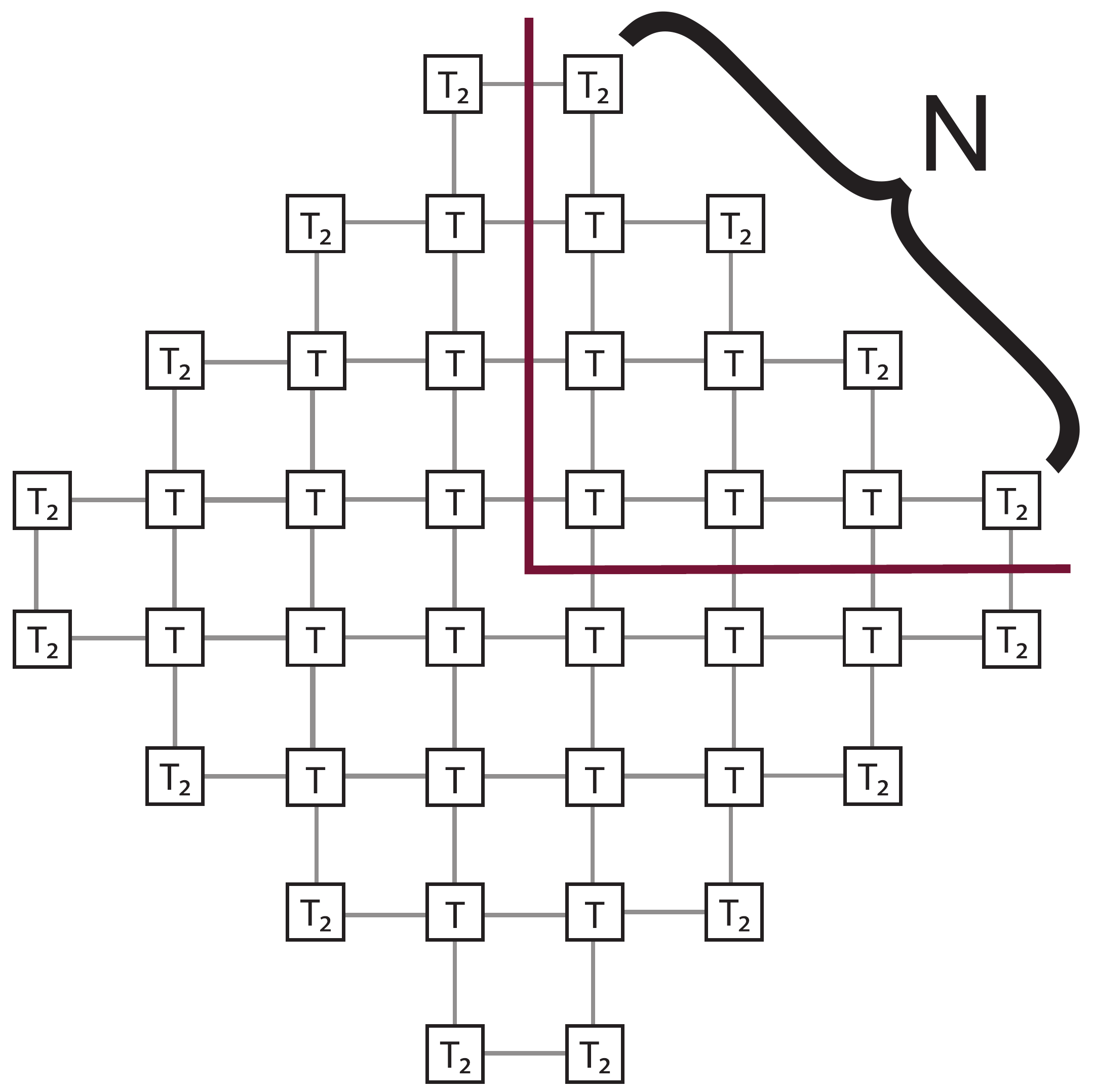}
\caption[2D tensor network that is to be contracted.]{The choice of the diamond-shape for the lattice $\Lambda$ is naturally suited to the application of corner transfer matrices to tensor networks. The tensor $T$ is defined by equation \eqref{eq:tensorT}, the tensor $T_2$ by equation \eqref{eq:tensorT2}. The number of tensors per lattice side is denoted by $N$ so that $\left|\Lambda\right|=N^2$. The tensors are interconnected with $2^k$-leveled spins. The violet lines isolate a quarter of the lattice to be used for CTM.}
\label{fig:TensCirc1}
\end{figure}

As will become clear soon, it is beneficial to take $\Lambda$ to be a diamond-shaped lattice presented in figure \ref{fig:TensCirc1}. Also, open boundary conditions are the natural choice for the method.\footnote{Versions of CTM that would allow for periodic boundary conditions are suggested in \cite{Orus2012,Bartel2008} but they are claimed to be much more computationally demanding than CTM with open boundary conditions.} Each interior lattice site is occupied by a tensor $T$ defined by equation \eqref{eq:tensorT}. The boundary sites are occupied by tensors $T_2$ defined as:
\begin{equation}
T_{2}^{\left(k\right)}\left(\underline{\alpha},\underline{\beta}\right):=\frac{1}{2}\text{tr}\left(\prod_{j=1}^{k}o^{\alpha_{j}}o^{\beta_{j}}\,a\right).\label{eq:tensorT2}
\end{equation}
They are functions of just two neighboring $2^k$-valued spins, i.e., they are rank-2 tensors. Again, the factor $\frac{1}{2}$ is to account for overall normalization. We can denote the number of the tensors (which is equal to the number of contracted physical spins) per side of the lattice by $N$ so that $\left|\Lambda\right|=N^2$. The tensors are interconnected by $2^k$-leveled spins, the merged ancillary spins.

\begin{figure}[H]
	\centering
	\hspace{1.8cm}\includegraphics[width=0.81\linewidth]{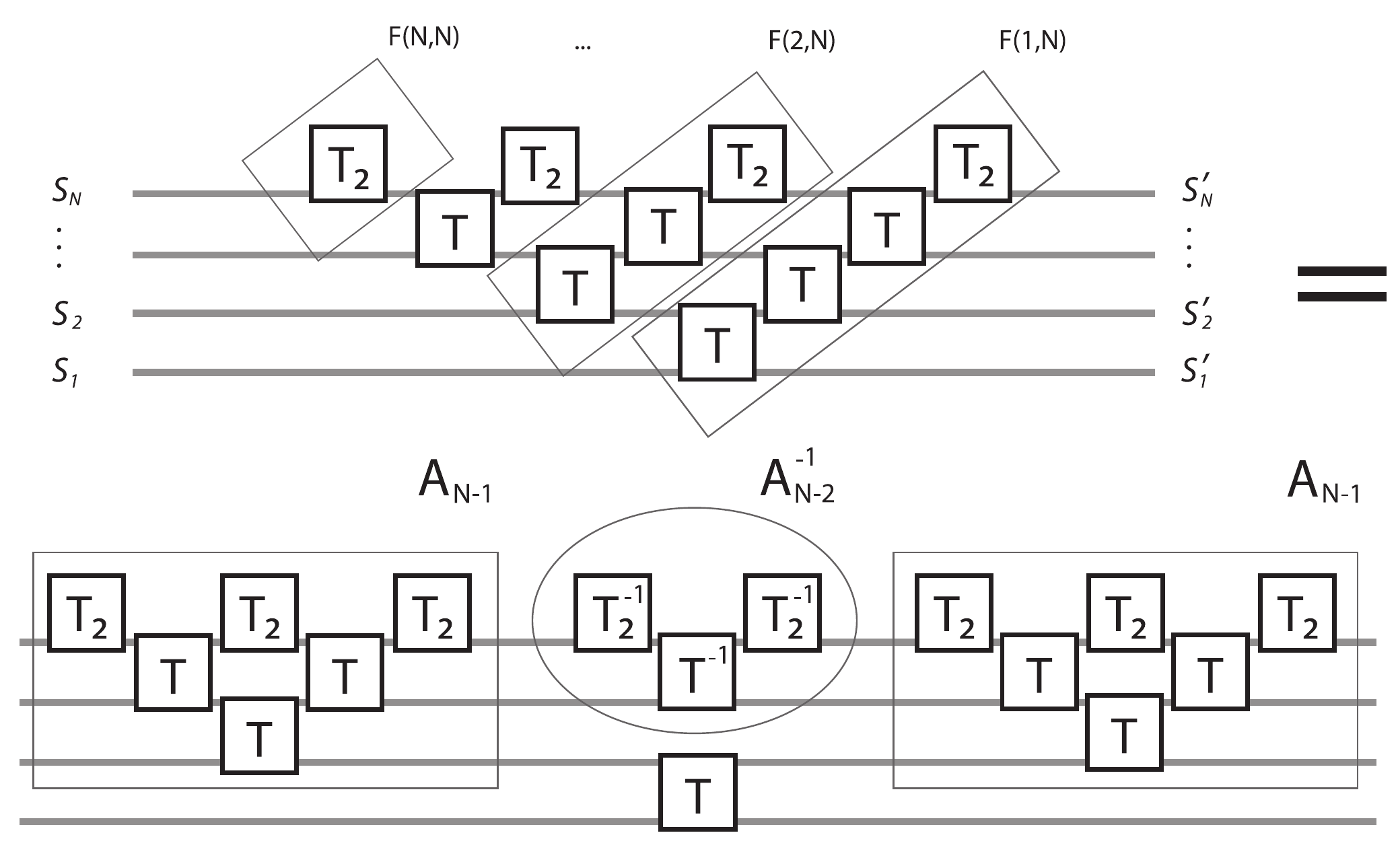}
	\caption[Quantum circuit identity that illustrates the recursion relation.]{A quarter of the tensor network from figure \ref{fig:TensCirc1} can be represented as the quantum circuit in the upper part of this figure. In order to do that, the elements of $T$ and $T_2$ have to be arranged in a $2^{2k}\times2^{2k}$ matrix $T\left(\underline{\alpha},\underline{\beta}|\underline{\gamma},\underline{\delta}\right)$ and a $2^{k}\times2^{k}$ matrix $T_2\left(\underline{\alpha}|\underline{\beta}\right)$. The horizontal lines represent the $2^k$-valued spins. The matrices $F\left(1,N\right)$, ..., $F\left(N,N\right)$ are marked to illustrate the definition \eqref{eq:AwithFs}. The figure as a whole is a representation of the CTM recursion relation \eqref{eq:RecursionCTM} as a quantum circuit identity. The matrix $A_{N}$ (on top) can be written in terms of two matrices $A_{N-1}^{*}$ (bottom left and right), a matrix $\left(A_{N-2}^{**}\right)^{-1}$ (bottom center top) and a matrix $T\left(1,N\right)$ (bottom center bottom). For notation see the main text.}
	\label{fig:TensCirc2}
\end{figure}

To apply CTM, one has to divide the lattice into quarters as denoted in figure \ref{fig:TensCirc1}. A quarter (or a corner) can be represented as the quantum circuit in the upper part of figure \ref{fig:TensCirc2}. Here $T^{\left(k\right)}$ has to be arranged in a $2^{2k}\times2^{2k}$ matrix $T^{\left(k\right)}\left(\underline{\alpha},\underline{\beta}|\underline{\gamma},\underline{\delta}\right)$ and $T_2^{\left(k\right)}$ in a $2^{k}\times2^{k}$ matrix $T_2^{\left(k\right)}\left(\underline{\alpha}|\underline{\beta}\right)$. From now on we will omit the arguments [for example 
$\left(\underline{\alpha},\underline{\beta}|\underline{\gamma},\underline{\delta}\right)$] and treat $T^{\left(k\right)}$ and $T_2^{\left(k\right)}$ simply as matrices. We shall closely follow the standard CTM derivation \cite{BaxterESM,BaxterCTM1981}, however, adopting notation to tensor circuits. We can denote the matrix $T^{\left(k\right)}$ acting on the $i$-th and $\left(i+1\right)$-th 
auxiliary spin of a $N$-spin circuit (with the spins numbered from bottom up in the circuits) by:
\begin{equation}
	T^{\left(k\right)}\left(i,N\right):=\underbrace{I\otimes\ldots\otimes I}_{i-1}\otimes T^{\left(k\right)}\otimes\underbrace{I\otimes\ldots\otimes I}_{N-i-1}, \label{eq:T(i,N)}
\end{equation}
where $I$ is the $2^k\times 2^k$ identity matrix. The tensor $T_2^{\left(k\right)}$ always acts on the last spin:
\begin{equation}
	T_2^{\left(k\right)}\left(N,N\right):=\underbrace{I\otimes\ldots\otimes I}_{N-1}\otimes T_2^{\left(k\right)}.
\end{equation}

We define the \textit{corner transfer matrix} $A^{(k)}_{N}$ to represent the quarter of the lattice:
\begin{equation}
	A^{(k)}_N:=F^{\left(k\right)}\left(1,N\right)F^{\left(k\right)}\left(2,N\right)\cdots F^{\left(k\right)}\left(N,N\right),\label{eq:AwithFs}
\end{equation}
with
\begin{equation}
	F^{\left(k\right)}\left(i,N\right):=T_2^{\left(k\right)}\left(N,N\right)T^{\left(k\right)}\left(N-1,N\right)\cdots T^{\left(k\right)}\left(i,N\right). \label{eq:MatrixF}
\end{equation}
To give the intuition, the definition \eqref{eq:AwithFs} is visualized in figure \ref{fig:TensCirc2}. The superscript index $k$ is sometimes omitted for brevity when it is clear from the context. The matrix $A^{(k)}_N$ is a $2^{kN}\times2^{kN}$ matrix. It acts on the incoming auxiliary spins $\underline{s}$ and transfers their values over the corner of the lattice to the outgoing auxiliary spins $\underline{s}'$. This is where the name corner transfer matrix originates. The matrix $A^{(k)}_N$ is symmetric as the circuit in figure \ref{fig:TensCirc2} is invariant under mirroring around the vertical axis (that is $\underline{s}\leftrightarrow\underline{s}'$). Equation \eqref{eq:FourCoef} for $\rho_{k}$ can be expressed in terms of a product of four such matrices:
\begin{equation}
\rho_{k}\left(N\right)=\text{tr}\left[\left(A^{(k)}_N\right)^{4}\right]. \label{eq:A^4}
\end{equation}
As explained, the normalization factor $\frac{1}{\mathcal{N}}$ from equation \eqref{eq:FourCoef} has been accounted for by the divisions by 2 in the definitions of tensors $T$, equations (\ref{eq:tensorT},\ref{eq:tensorT2}). The contraction of the 3D tensor network from figure \ref{fig:3DTensNet} has therefore been reduced to evaluating the trace of the corner transfer matrix to the fourth power. However, we still need an efficient way to compute $A^{(k)}_N$.

By introducing further notation:
\begin{eqnarray}
B^*&:=&I\otimes B,\label{eq:Star} \\
B^{**}&:=&I\otimes I\otimes B,\label{eq:twoStars}
\end{eqnarray}
equation \eqref{eq:AwithFs} can be rewritten as the following recursion relation:
\begin{equation}
	A_{N}=F\left(1,N\right)A_{N-1}^{*},\label{eq:RecursionSmall}
\end{equation}
Equation \eqref{eq:RecursionSmall} can be further simplified (writing it for $A_{N-1}^{*}$ and using $F\left(1,N\right)=F\left(2,N\right)T\left(1,N\right)$) to get the standard CTM recursion relation:
\begin{equation}
	A_{N}=A_{N-1}^{*}\left(A_{N-2}^{**}\right)^{-1}T\left(1,N\right)A_{N-1}^{*}.\label{eq:RecursionCTM}
\end{equation}
It links the matrix $A_{N}$ from the $N$-th step of iteration (corresponding to the system of $N\times N$ spins) to the matrices $A_{N-1}$ and $A_{N-2}$ from the previous two steps of iteration (corresponding to the systems of $\left(N-1\right)\times \left(N-1\right)$ and $\left(N-2\right)\times \left(N-2\right)$ spins, respectively). This recursion relation can be nicely illustrated by the quantum circuit identity in figure \ref{fig:TensCirc2}.

\subsection{Renormalized recursion} \label{sec:RenormRecur}
Recursion relation \eqref{eq:RecursionCTM} is just a step away from enabling iterative computation of $\rho_k\left(N\right)$ for large $N$: the size of the matrices in the formula still grows as $2^{kN}\times2^{kN}$ so the algorithm unavoidably exhausts the available memory space at some relatively small $N$. As usual in CTM method, this is overcome by diagonalization of the $A$'s:
\begin{eqnarray}
A_{N}&=&\alpha_{N}P_{N}A_{N}^{d}P_{N}^{\top},\label{eq:DiagA}\\
P_{N}^{\top}P_{N}&=&I.\label{eq:OrthoP}
\end{eqnarray}
Orthogonality of the diagonalization is achieved because of symmetricity of $A_{N}$. The real eigenvalues of $A_{N}$ are sorted in order of decreasing magnitude and normalized by the largest eigenvalue $\alpha_{N}$ so that the largest element of $A_{N}^{d}$ is 1.

We can then rewrite relation \eqref{eq:RecursionCTM} as:
\begin{eqnarray}
\mathcal{A}_{N}&:=&A_{N-1}^{d*}\left(R_{N-1}^{*}\right)^{\top}\left(A_{N-2}^{d**}\right)^{-1}T\left(1,N\right)R_{N-1}^{*}A_{N-1}^{d*},\label{eq:RecursionDiag}\\
\mathcal{A}_{N}&=&\kappa_{N}R_{N}A_{N}^{d}R_{N}^{\top},\label{eq:DiagMathA}
\end{eqnarray}
where the relation between $\mathcal{A}$, $R$, $\kappa$ and $A$, $P$, $\alpha$ is given with equations (\ref{eq:DiagA}-\ref{eq:DiagMathA}) and:
\begin{eqnarray}
R_{N}&:=&\left(P_{N-1}^{*}\right)^{\top}P_{N},\label{eq:MatixR}\\
\kappa_{N}&:=&\alpha_{N}\alpha_{N-2}/\alpha_{N-1}^{2}. \label{eq:kappa}
\end{eqnarray}
Here we have used the fact that $\left(P_{N-2}^{**}\right)^{\top}$ and $T\left(1,N\right)$ commute, which follows from their structures (equation \eqref{eq:twoStars} and equation \eqref{eq:T(i,N)} for $i=1$). From equations \eqref{eq:MatixR} and \eqref{eq:OrthoP} it follows that $R_N$ is orthogonal, as well.

Recursion relations (\ref{eq:RecursionDiag}-\ref{eq:kappa}) are written in the form that allows for renormalization - truncation of small eigenvalues in every step of iteration by which we keep the sizes of the matrices limited despite increasing $N$. This is not only due to using the diagonal form the matrices. For example, if we truncated $A_{N}^{d}$ in equation \eqref{eq:DiagA} and the corresponding columns of $P_N$, the number of rows of $P_N$ would remain unchanged and would still blow up with $N$. With definition \eqref{eq:MatixR}, the inflating dimensions of $P_{N}$ and $\left(P_{N-1}^{*}\right)^{\top}$ are hidden in the product so that with truncation the size of $R_{N}$ remains limited when increasing $N$. Of course, after the first truncation, the $A_N$ and $P_N$ cannot be reconstructed anymore and we are left with the composed matrices $\mathcal{A}_{N}$ and $R_{N}$.

We implemented the algorithm as follows. First we prepare, using recursion relation \eqref{eq:RecursionSmall}, the matrices $A_N$ for $N=1,\ldots,N_0$ for a chosen small $N_0$ (typically between 3 and 10). Using equation \eqref{eq:DiagA} we diagonalize $A_{N_0}$ and $A_{N_0-1}$ and compute $R_{N_0}$ using equation \eqref{eq:MatixR}. Up to this point the computation is exact. We continue the iteration using recursion formulae (\ref{eq:RecursionDiag},\ref{eq:DiagMathA}) and renormalize the matrices in every step. Also, from this point on, the matrix $T$ is not increasing in size any more - we use $T\left(1,N_0+1\right)$ and truncate it in every step of iteration as will be explained shortly. This sets the limit to the maximal size of any matrix in the algorithm to $2^{k\left(N_0+1\right)}\times2^{k\left(N_0+1\right)}$.

Truncation is done by cutting away the (normalized) eigenvalues of $A_{N}^{d}$ below the chosen parameter $\epsilon$. As the maximal size of matrices in the algorithm is limited, we have a further constraint that no more than $2^{k\left(N_0-1\right)}$ eigenvalues can be kept (so that the product $\left(A_{N}^{d**}\right)^{-1}T\left(1,N_0+1\right)$ can be defined two steps of iteration later). Sometimes, when calculating for larger number $k$ of time steps, it is useful to take this number to be even smaller, say $c\cdot 2^{k\left(N_0-1\right)}$ for $0<c\leq 1$, to reduce the memory consumption. After the truncation of $A_{N}^{d}$ also the corresponding columns in $R_{N}$ have to be truncated so that the product $R_{N}A_{N}^{d}$ is well defined. If $A_{N-2}^{d}$ has been truncated to a smaller number of diagonal elements than $2^{k\left(N_0-1\right)}$, the appropriate rows in $T\left(1,N_0+1\right)$ have to be truncated so that the product $\left(A_{N-2}^{d**}\right)^{-1}T\left(1,N_0+1\right)$ is well defined - we have to truncate those rows of $T\left(1,N_0+1\right)$ that would be multiplied with the missing columns in $\left(A_{N-2}^{d**}\right)^{-1}$. To preserve commutativity of $\left(A_{N-2}^{d**}\right)^{-1}$ and $T\left(1,N_0+1\right)$, we also truncate the corresponding columns of $T\left(1,N_0+1\right)$. In practice, $T\left(1,N_0+1\right)$ from definition \eqref{eq:T(i,N)} is stored in a dedicated constant and truncated accordingly in every iteration step. It is the easiest to see that no information is lost from $T\left(1,N_0+1\right)$ with truncation by taking a look at the structures of $\left(A_{N-2}^{d**}\right)^{-1}$ and $T\left(1,N_0+1\right)$:
\begin{eqnarray}
\left(A_{N-2}^{d**}\right)^{-1}&=&I\otimes I\otimes \left(A_{N-2}^{d}\right)^{-1},\\
T\left(1,N_0+1\right)&=&T\otimes \underbrace{I\otimes\ldots\otimes I}_{N_0-1}.
\end{eqnarray}
When truncating $A_{N-2}^{-1}$ and truncating $T\left(1,N_0+1\right)$ to fit $\left(A_{N-2}^{d**}\right)^{-1}$  we are manipulating just the identity part of $T\left(1,N_0+1\right)$ so really, no information is lost.

\begin{figure}[H]
	\centering
	\includegraphics[width=0.8\linewidth]{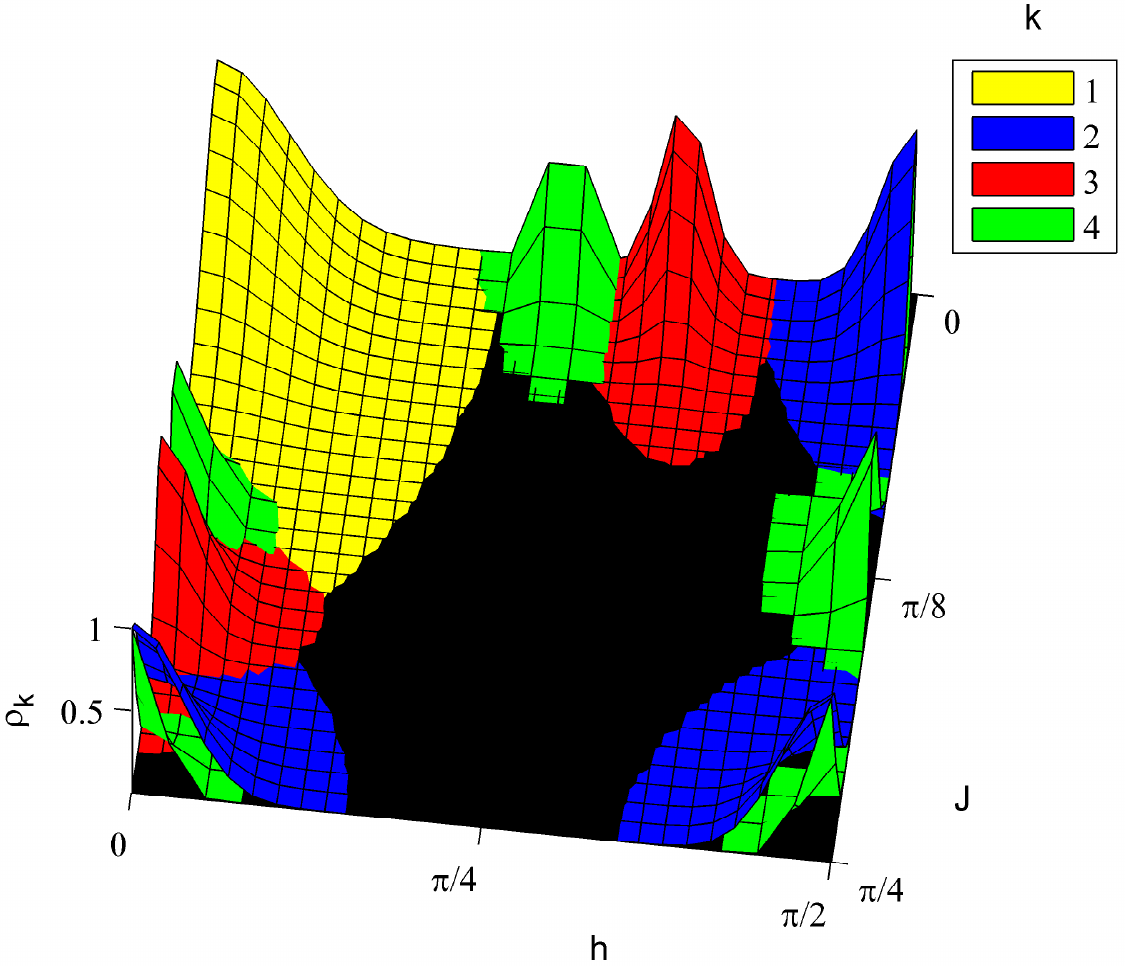}
	\vspace{0.3cm}
	
	\centering
	\includegraphics[width=0.85\linewidth]{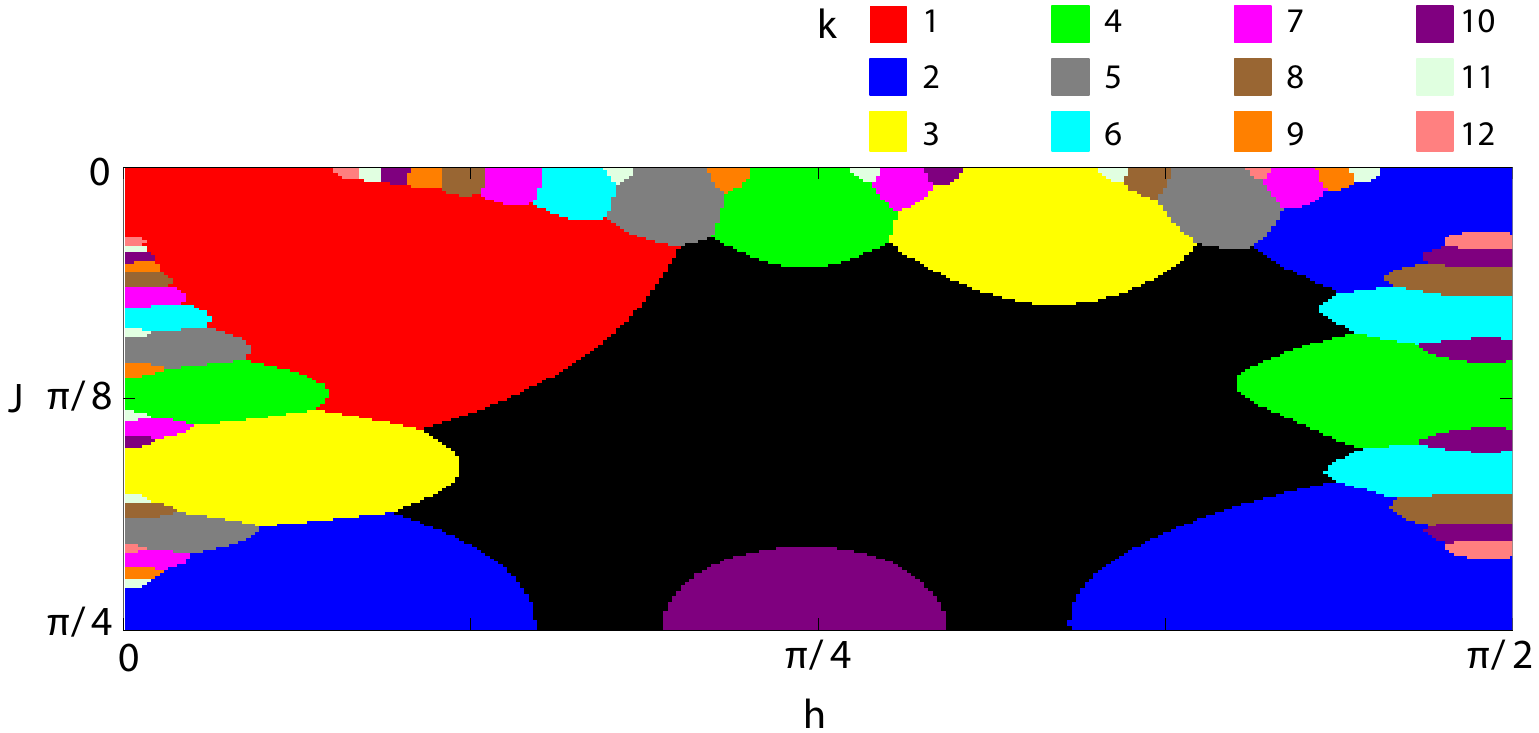}
	\caption[Phase diagram for small latices.]{The phase diagram $\rho_k\left(h,J\right)$ at $N=4$ (for the $4\times4$ lattice) for $k=1,\ldots,4$ obtained with our method based on CTM (top panel). The parameter space splits into regions where single Fourier modes are dominating. The modes are presented in colors as denoted in the legend. In the central region which is colored black all the modes are below the noise $\rho_\text{noise}\propto2^{-N^2/2}$. The result matches perfectly the diagram obtained in Ref.~\cite{PinedaProsen2014} (bottom panel). 
	This evokes trust in the developed method but is surprising at the same time. Although periodic boundary conditions were used in Ref.~\cite{PinedaProsen2014} while here 
	we are using open boundary conditions we still find quite remarkable quantitative agreement.}
	\label{fig:SmallLattice}
\end{figure}

We run the iteration up to the desired $N_{\text{max}}$. In every step we can evaluate $\rho_{k}\left(N\right)$. Equation \eqref{eq:A^4} in terms of diagonalized and truncated corner transfer matrices is written as:
\begin{equation}
\rho_{k}\left(N\right)= [\alpha^{(k)}_{N}]^{4}\,\text{tr}\left[A^{(k)d}_N\right]^{4}. \label{eq:Contraction}
\end{equation}
The number $\alpha_{N}^{(k)}$ is computed from $\kappa_{N}^{(k)}$, $\alpha_{N-1}^{(k)}$ and $\alpha_{N-2}^{(k)}$ by inversion of equation \eqref{eq:kappa}. In practice, the Fourier coefficient $\rho_k\left(N\right)$ for larger $N$ falls bellow the smallest positive number supported by the programming language and cannot be distinguished from zero anymore. On the contrary, $\kappa_N^{(k)}$ remains of the same order of magnitude for all $N$. It therefore makes sense to use logarithmized equation \eqref{eq:Contraction} (and logarithmized inverted equation \eqref{eq:kappa}) to overcome this inconvenience and be able to reach arbitrarily large $N$.

We can use the cumulative sum of the absolute values of the truncated eigenvalues of $\mathcal{A}$'s (sum of all the truncated eigenvalues up to the $N$-th step of iteration) to estimate the relative error of $\alpha_N$. The eigenvalues had been renormalized in equation \eqref{eq:DiagMathA} so they can serve as an estimate for the {\em relative} error. The \textit{relative error of} $\rho_k\left(N\right)$, $\delta\left[\rho_k\left(N\right)\right]$, can then be taken to be four times this cumulative sum ($\alpha^{(k)}_N$ is to the power of 4 in equation \eqref{eq:Contraction}). This is probably a considerable overestimation of the actual error because the computed value $\rho_k\left(N\right)$ most probably does not drift away sistematically from the exact value but rather oscillates around it. Therefore, this serves only as a rough estimate of the quality of the algorithm.

This completes our renormalizable method that enables numerically exact computation of the level density of the quasienergy spectra of 2D kicked quantum systems.

\section{Results}
\label{sec:Results}

The main advantage of the presented numerical technique is its renormalizability. It enables computation of the properties of systems of very large lattice sizes within a polynomial time. However, the method also has+ two shortcomings: First, because for a given $k$ the size of the largest matrices that appear in the algorithm  is 
$2^{k\left(N_0+1\right)}\times2^{k\left(N_0+1\right)}$ for a chosen value of the parameter $N_0\geq3$, we are limited to using the method for short propagation times $k$. 
In case of 2D KI model, we were able to treat $k=1,2,3,4$ using our computer cluster. 
Second, it turns out as will be explained in details later, that the method is renormalizable, i.e. gives convergent results in the thermodynamic limit, only in certain parts of the parameter space.

\begin{figure}[H]
\centering
\includegraphics[width=1\linewidth]{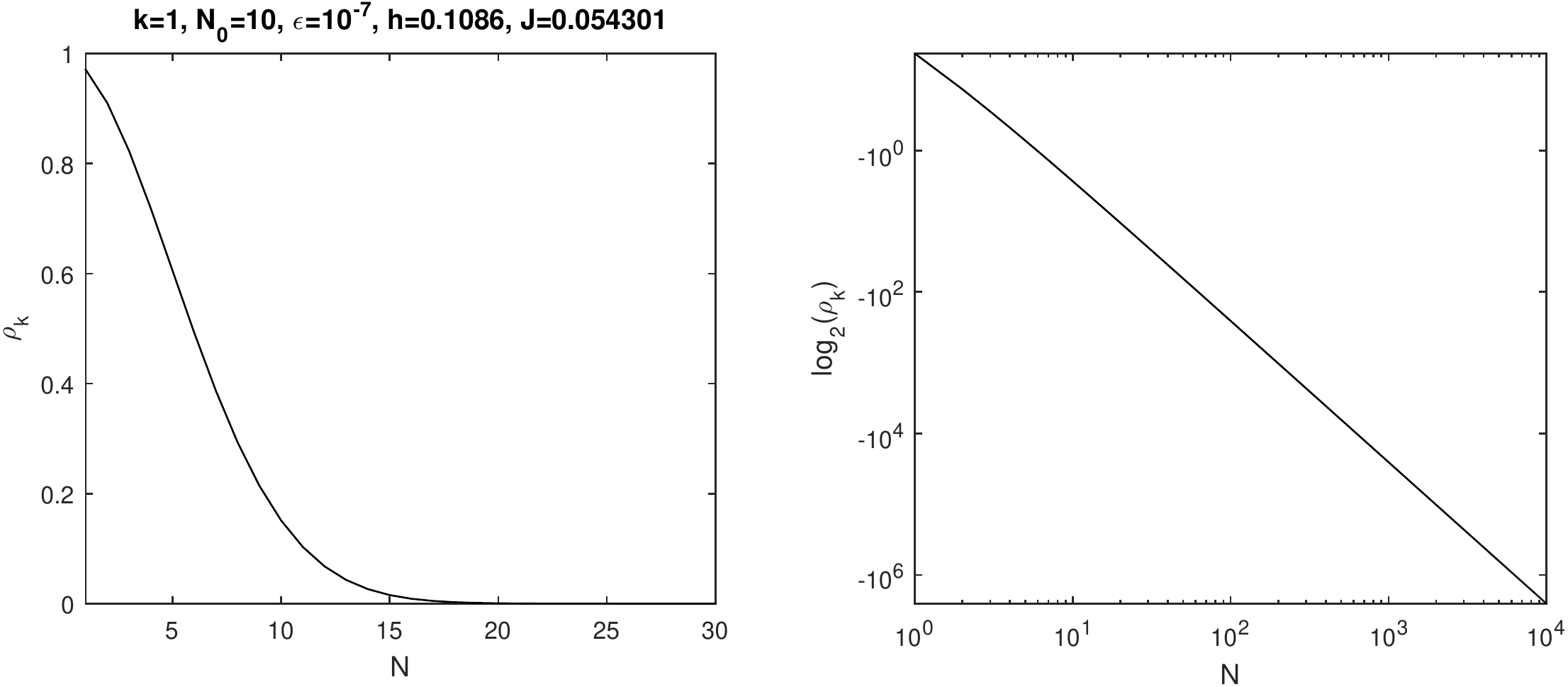}
\caption[Typical dependence $\rho_k\left(N\right)$]{Typical dependence of the Fourier modes $\rho_k\left(N\right)$. The particular parameters used here are listed in the figure, however the dependence is qualitatively the same for any values of the parameters where the method is renormalizable. The left diagram is a close-up, the right diagram shows $\log_2\left(\rho_k\left(N\right)\right)$ in log-log scale which makes the comparison with the prediction from random matrix theory (equation \eqref{eq:RMTprediction}) easier. As predicted, $\rho_k\left(N\right)$ decays exponentially with $N^2$.}
\label{fig:TypicalDependence}
\end{figure}

We have implemented the method in Matlab and computed the dependencies $\rho_{k}\left(N\right)$ for different values of the system's parameters $h$ and $J$ 
(for transversal exterior magnetic field, $\theta=0$). As explained in Ref.~\cite{PinedaProsen2014}, because of the symmetries of the Hamiltonian, the spectrum of the Floquet map is $\pi$-periodic in $h$, $\pi/2$-periodic in $J$ and invariant to the change of sign of $h$ or $J$. It therefore suffices to study it on the domain $\left(h,J\right)\in\left[0,\pi/2\right]\times\left[0,\pi/4\right]$. We used uniform discretization of this domain with different resolutions. In general, for each point of the parameter space, $\rho_{k}\left(N\right)$ was computed up to $N_{\text{max}}=10\,000$ (corresponding to a $10\,000\times10\,000$ lattice) but with an additional time-of-computation limit to keep the overall computation time under control. The time limit was long enough so that $N_{\text{max}}$ was reached for all the points of the parameter space where the algorithm was renormalizable. For every $k$, the largest value of the parameter $N_0$ that still allowed for efficient run times on our computer cluster was chosen. As will be demonstrated, the algorithm is robust to changing the value of the truncation parameter $\epsilon$ over a couple of decades. We used $\epsilon$ from the middle of this interval.

In the following, the main final results will be presented first. After that we will elaborate on more detailed results and the relevant characteristics of the method.

\subsection{Main results}

To see if the developed method could be expected to give reasonable results, we compared the phase diagram $\rho_k\left(h,J\right)$ for small lattices with the interesting phase diagram from \cite{PinedaProsen2014}. The comparison is presented in figure \ref{fig:SmallLattice}. The two diagrams match perfectly which evokes trust in the algorithm. At the same time, the matching is also slightly surprising because of the different boundary conditions that were used.

\begin{figure}[H]
	\centering
	\includegraphics[width=0.495\linewidth]{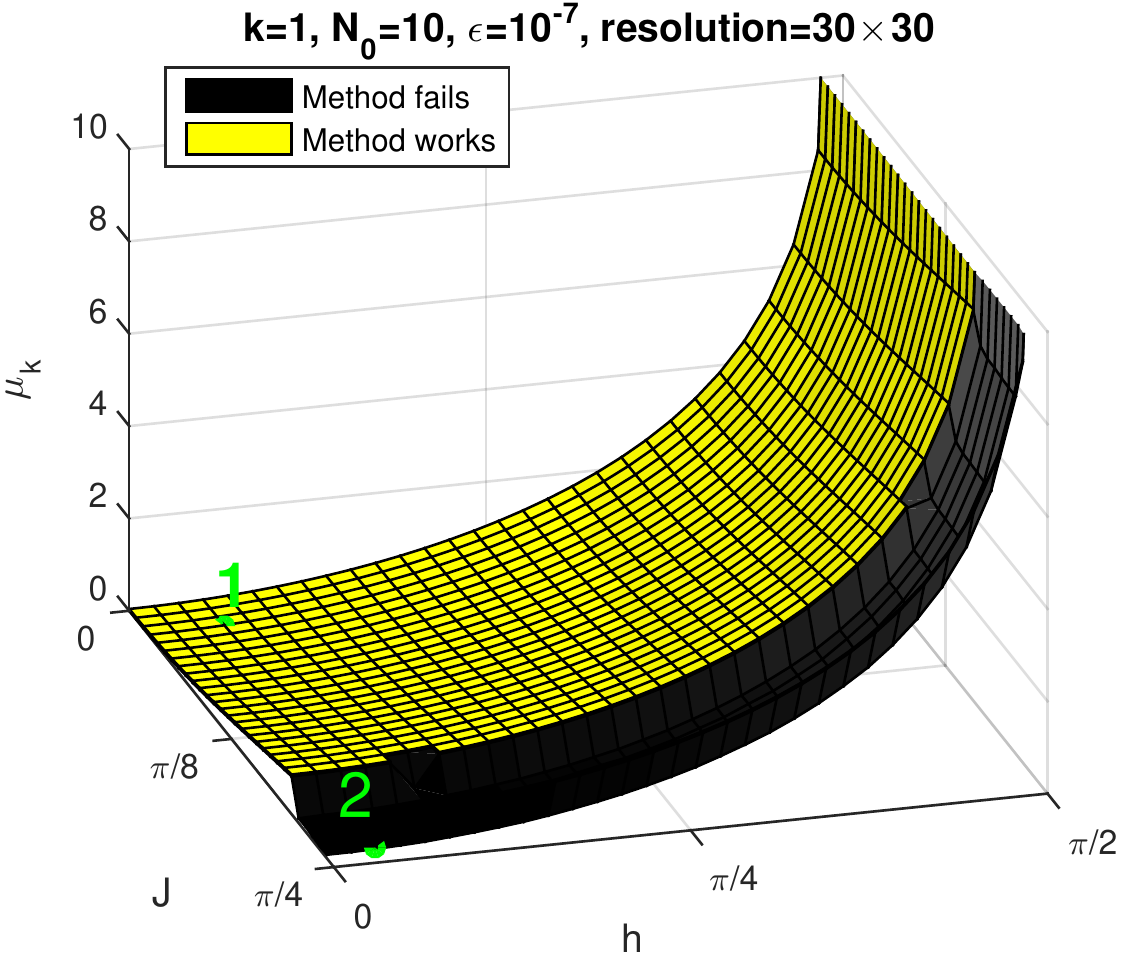}
	\includegraphics[width=0.495\linewidth]{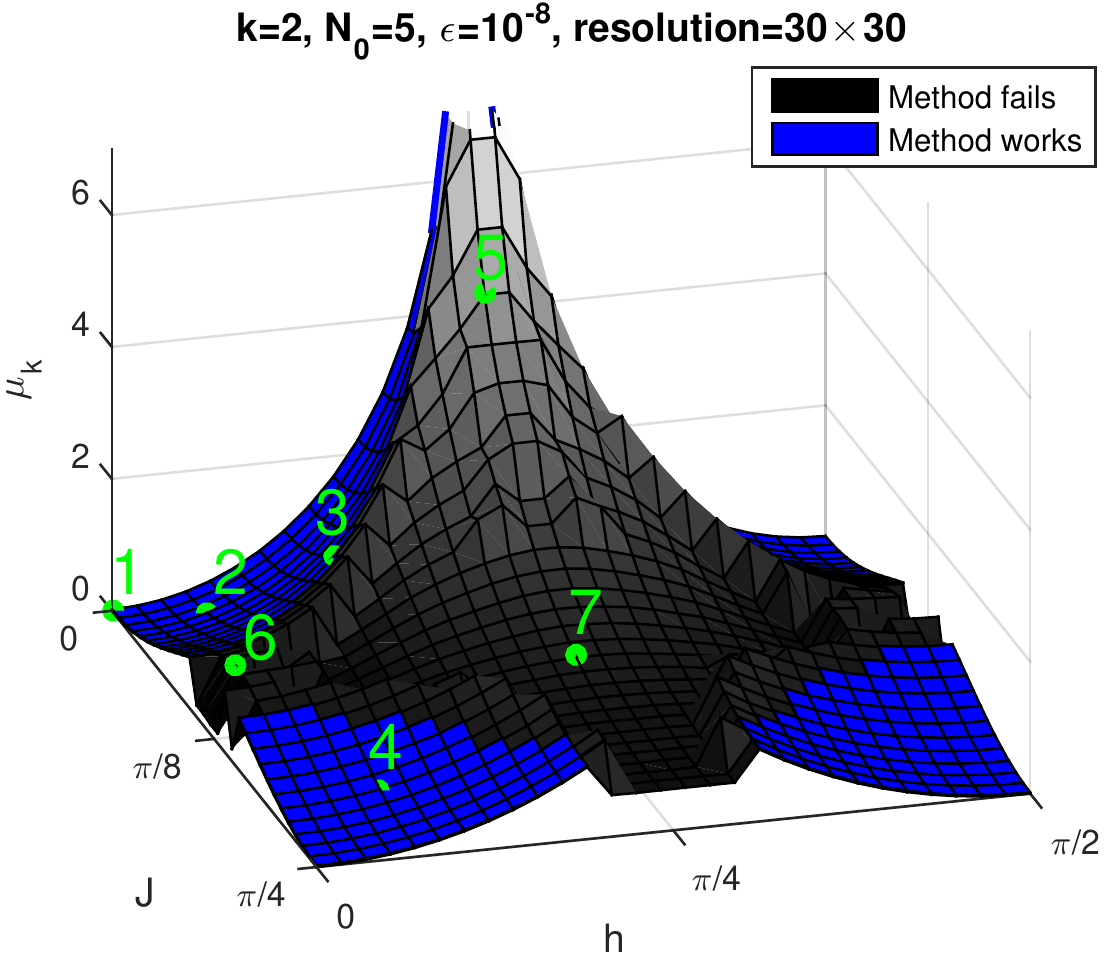} \\
	\includegraphics[width=0.48\linewidth]{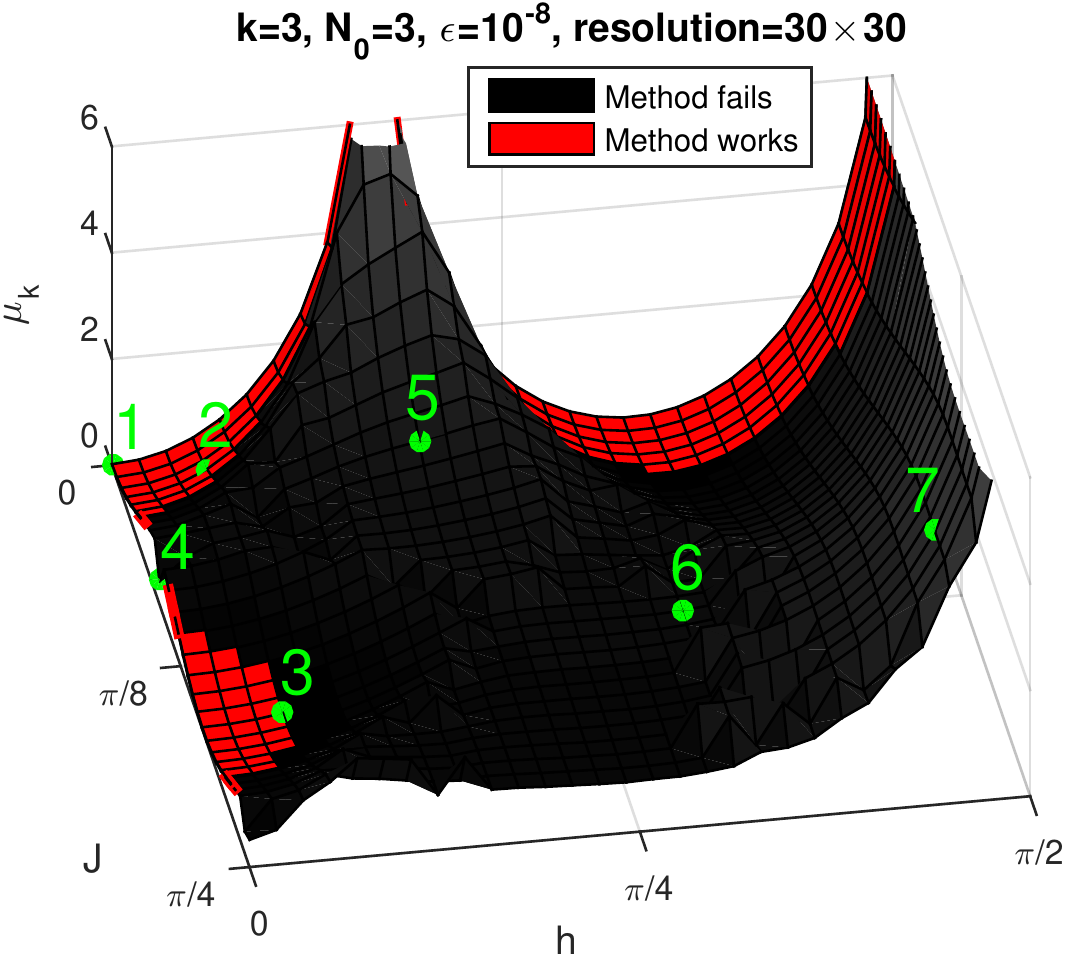}
	\includegraphics[width=0.50\linewidth]{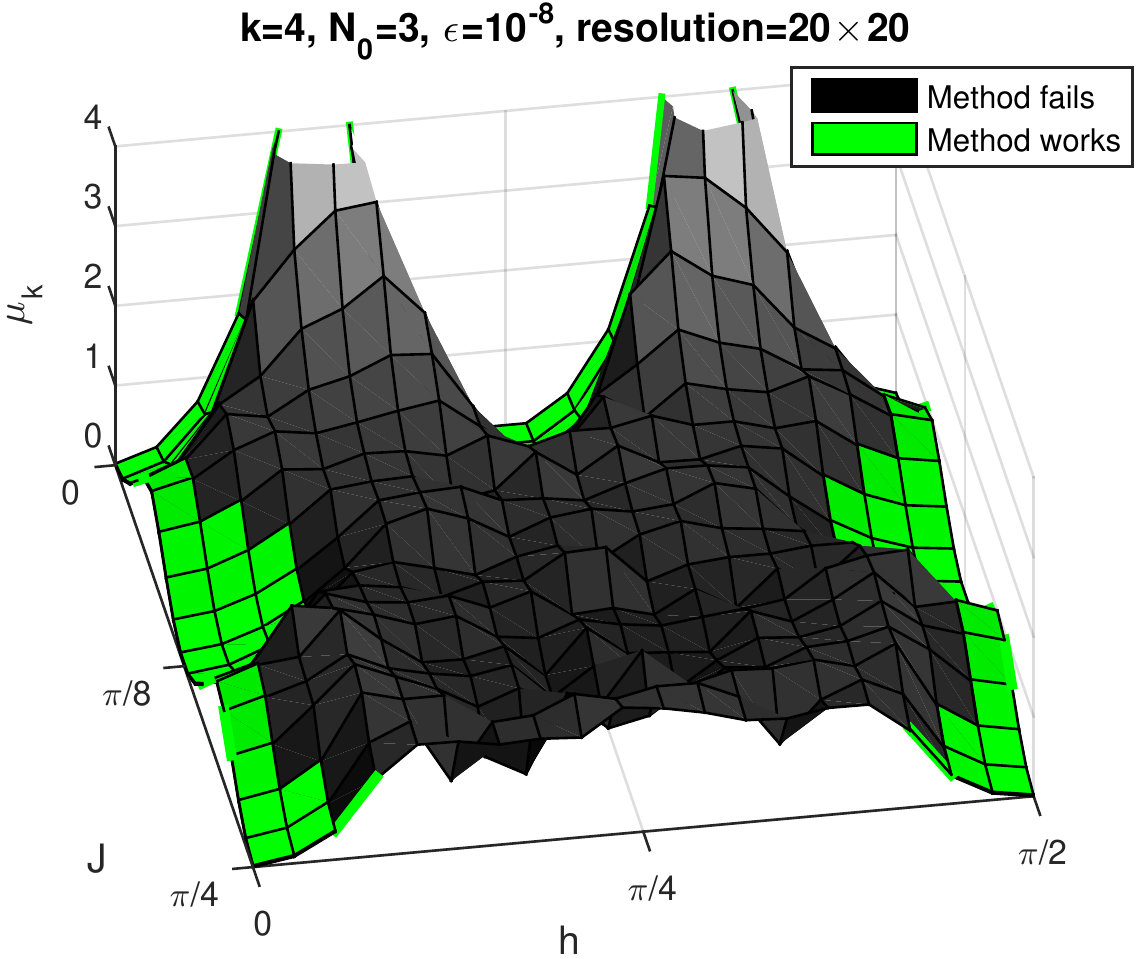}
	\caption[Phase diagram $\mu_k\left(h,J\right)$ for $k=1,2,3,4$]{Phase diagrams $\mu_k\left(h,J\right)$ (as defined by equation \eqref{eq:Model} for the asymptotic behavior $N\rightarrow\infty$) for $k=1,2,3,4$. The diagram for $k=4$ was computed at a lower resolution than the rest because of the large time consumption of the computation. Contrary to the RMT prediction (equation \eqref{eq:RMTprediction}), the dependence of $\mu_k$ on the system's parameters is nontrivial. Comparing the above diagrams to the diagram $\rho_k\left(h,J\right)$ for small lattices in figure \ref{fig:SmallLattice}, we see that $\mu_k\left(h,J\right)$ preserve the structure of their finite-lattice counterparts -- $\mu_k\left(h,J\right)$ tend to zero for the points where where $\rho_k\left(h,J\right)$ for small $N$ have peaks. This indicates that for these points the level density of the quasienergy spectrum may be non-flat even in the thermodynamic limit. The parameter space consists of regions where the method is well working (colored) and the region where the method fails (shaded in gray). These regions were determined as described in the main text. The diagrams for the coefficient of the subleading term, $\nu_k$, were very similar to those of $\mu_k$ so they are not presented. In general, in the regions where the method works, $\nu_k$ was limited as $\nu_k\left(h,J\right)\leq\text{const}\cdot\mu_k\left(h,J\right)$ with $\text{const}=1$ for $k=1,2$ and $\text{const}=2$ for $k=3,4$. The green numbers indicate representative points in the parameter space for which properties of the algorithm shall be analyzed later in the text.}
	\label{fig:Alpha(h,J)}
\end{figure}

As predicted with random matrix theory, $\rho_{k}\left(N\right)$ decays exponentially with $N$. Figure \ref{fig:TypicalDependence} shows the typical dependence obtained by CTM method for the 2D KI system. It can be therefore concluded  that the phases obtained for small lattices (figure \ref{fig:SmallLattice}) are, strictly speaking, only a finite-size effect. The scaling of $\rho_k\left(N\right)$ does not, however, quantitatively agree with the universal RMT prediction. First of all, we find that all data can be perfectly fitted with a model which includes the leading (volume) and subleading (surface) term in the exponential
\begin{equation}
\rho_{k}\left(N,h,J\right)=c_k\left(h,J\right)2^{-\mu_k\left(h,J\right) N^{2}-\nu_k\left(h,J\right) N}, \label{eq:Model}
\end{equation}
where the coefficients $c$, $\mu$ and $\nu$ are found to be nontrivially dependent on the system's parameters. Note that in a naive RMT model one would expect a universal result $\mu_k \equiv 1$ 
($\nu \equiv 0$; equation \eqref{eq:RMTprediction}).

Figure \ref{fig:Alpha(h,J)} shows the dependence of the leading coefficient $\mu_k$ from the model \eqref{eq:Model} on the parameters $h$ and $J$. We see that $\mu_k\left(h,J\right)$ is a nontrivial function. It is worth comparing these diagrams to the phase diagrams $\rho_k\left(h,J\right)$ for small lattices in figure \ref{fig:SmallLattice}. One notices that the diagrams $\mu_k\left(h,J\right)$ preserve the structure of their finite-lattice counterparts $\rho_k\left(h,J\right)$ -- in the regions where $\rho_k\left(h,J\right)$ have peaks, $\mu_k\left(h,J\right)$ generally tend to zero. This indicates that at the isolated ``peak points" the Fourier coefficients $\rho_k$ do not decay and the level density of the quasienergy spectrum stays non-flat even in the thermodynamic limit. This points are all confined to the boundary of the parameter space and correspond to rather degenerate limits of the 2D KI system.

What is also immediately noticed in the diagrams $\mu_k\left(h,J\right)$ is that the parameter space partitions into subsets where the CTM method works (i.e., provides convergent results in the thermodynamic limit) and subsets where it fails. This is probably one of the most interesting discoveries of our study. We determine whether the method works or not in a certain point using the relative error $\delta\left[\rho_k\left(N\right)\right]$ introduced in the previous section (\ref{sec:RenormRecur}). The criterium is that $N$ should be at least $N_c=1000$ before $\delta\left[\rho_k\left(N\right)\right]$ exceeds $\delta_c=0.1$. The numbers $N_c$ and $\delta_c$ are chosen rather arbitrarily so this is not to be regarded as the ultimate criterium for the functioning of the method. Especially, because the cumulative sum is probably a large overestimation of the actual error, as described earlier. However this gives us a visualization of what is happening and a way to eliminate evidently nonreliable points.

\begin{figure}[H]
\centering
\includegraphics[width=0.49\linewidth]{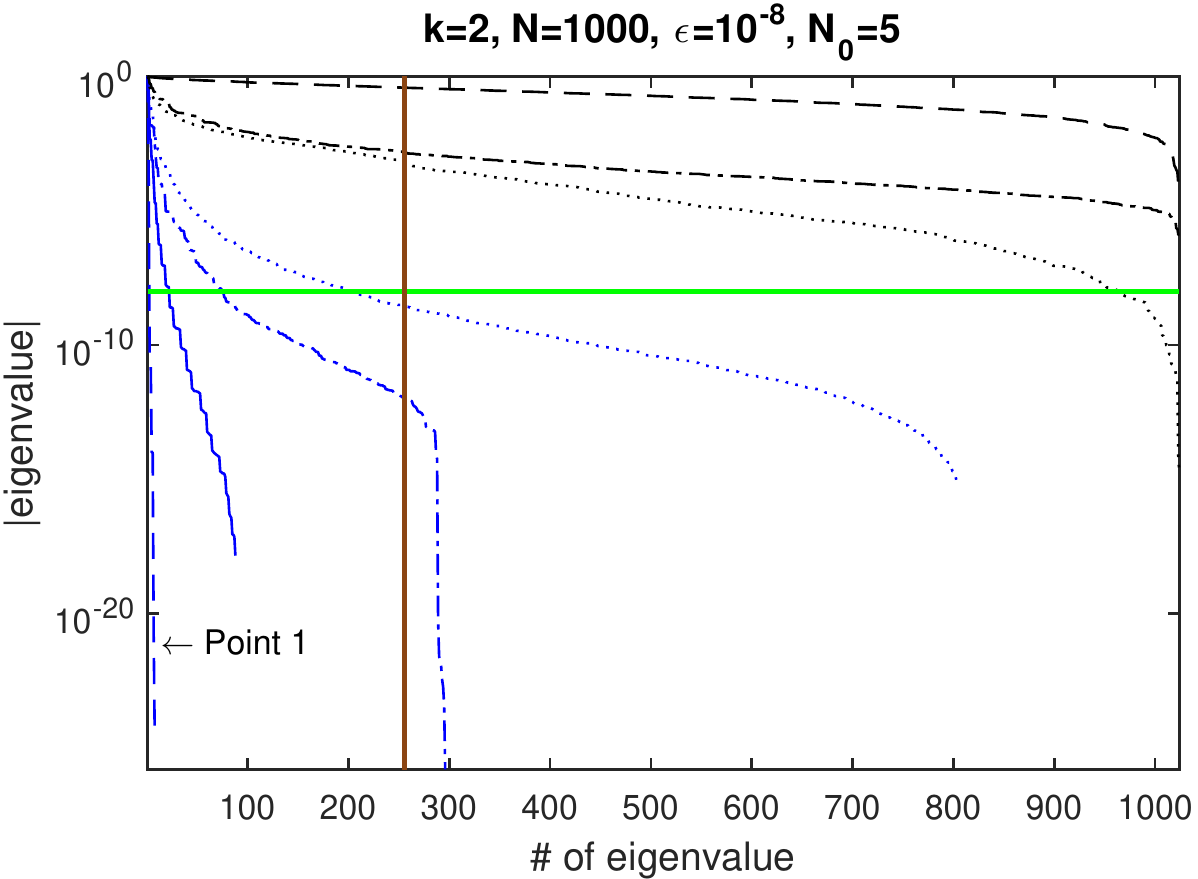}
\includegraphics[width=0.49\linewidth]{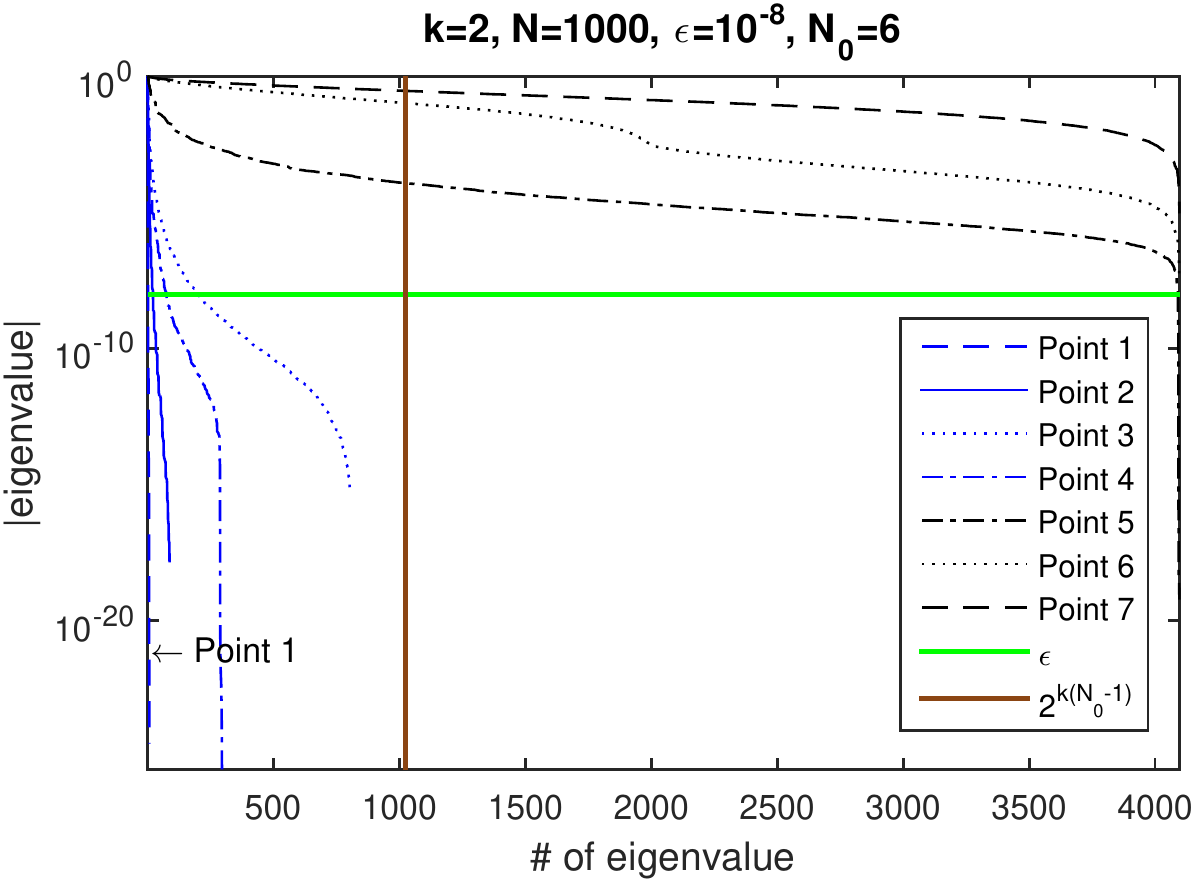}
\\
\includegraphics[width=0.49\linewidth]{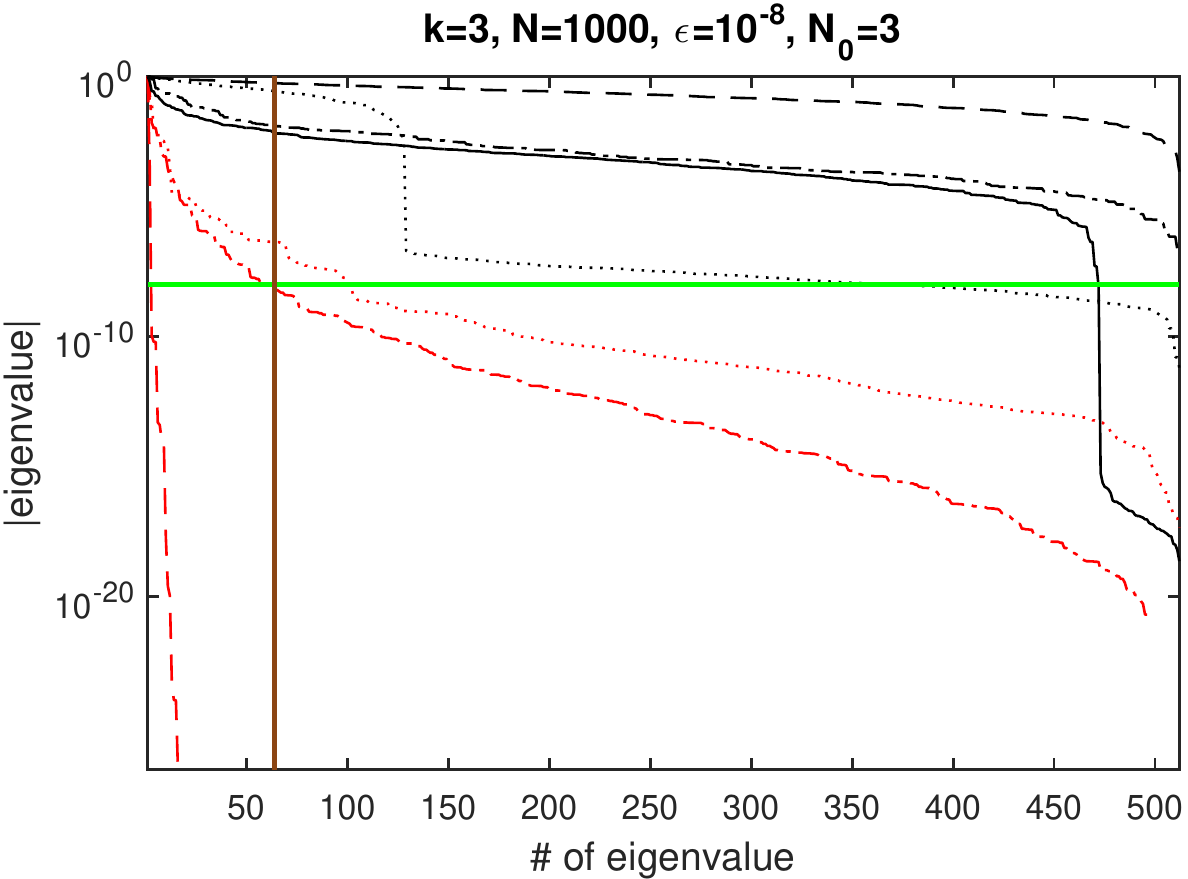}
\includegraphics[width=0.49\linewidth]{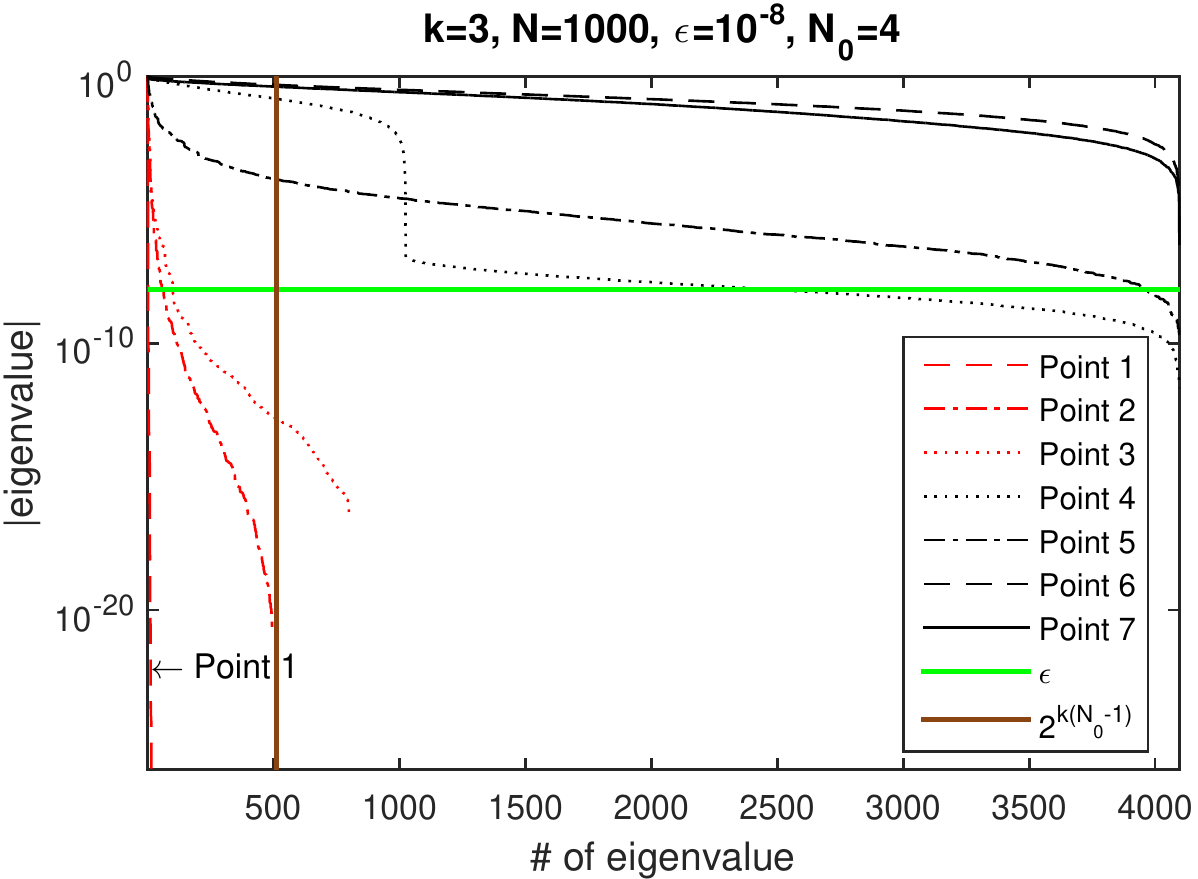}
\caption[Spectra of $A^{(k)d}_N$.]{Spectra of CTMs, $A^{(k)d}_N$, at $N=1000$ for $k=2,3$ for two different values of $N_0$ for the chosen set of typical points that is marked in the diagrams in figure \ref{fig:Alpha(h,J)}. The spectra at the ``working'' points are presented in colors and the spectra at the ``non-working" points are presented in black. When truncating the corner transfer matrices, we cut away all the eigenvalues smaller than $\epsilon$ (gren horizontal line) but with the additional requirement that no more than $2^{k\left(N_0-1\right)}$ eigenvalues are kept (brown vertical line). So, only the eigenvalues in the top left rectangle in each diagram are kept. The spectra at the ``working" points are steep enough that $\epsilon$ is determining the truncation (effectively finite rank of CTMs). The spectra at the ``non-working" points are flat (full rank of CTMs), so because of the eigenvalue number limit non-small eigenvalues are cut-away and the error grows rapidly. The method is non-renormalazible there. For some points at the border of the ``working" areas, for example Point 3 for $k=3$, $N_0=3$ the truncation can also be determined by the number limit but the spectra are steep enough so that the truncated eigenvalues are still small. The left two diagrams show the spectra at the values of $N_0$ as used for figure \ref{fig:Alpha(h,J)}. The data, in particular at points 4 and 5 for $k=3$, give the impression that the ``non-working" regions could be eliminated by increasing $N_0$. The right two diagrams show the spectra for $N_0$ increased by one, so the dimension of CTMs increased by a factor of $2^k$, comparing to the diagrams on the left. We see that the spectra of ``non-working" points just rescale and preserve the shape of the spectra at smaller $N_0$ and remain ``non-working". The spectra at the ``working" points are essentially unchanged and so appear steeper here due to the increased range of $x$-axis. The main difference is for the points that were close to the boundary of the ``working'' region, they fall within the ``working'' area more convincingly. Increasing $N_0$ may therefore enlarge the ``working'' regions slightly but at the cost of the exponentially increased computing resources.}
\label{fig:Spectra}
\end{figure}

\begin{figure}[H]
	\centering
	\includegraphics[width=0.61\linewidth]{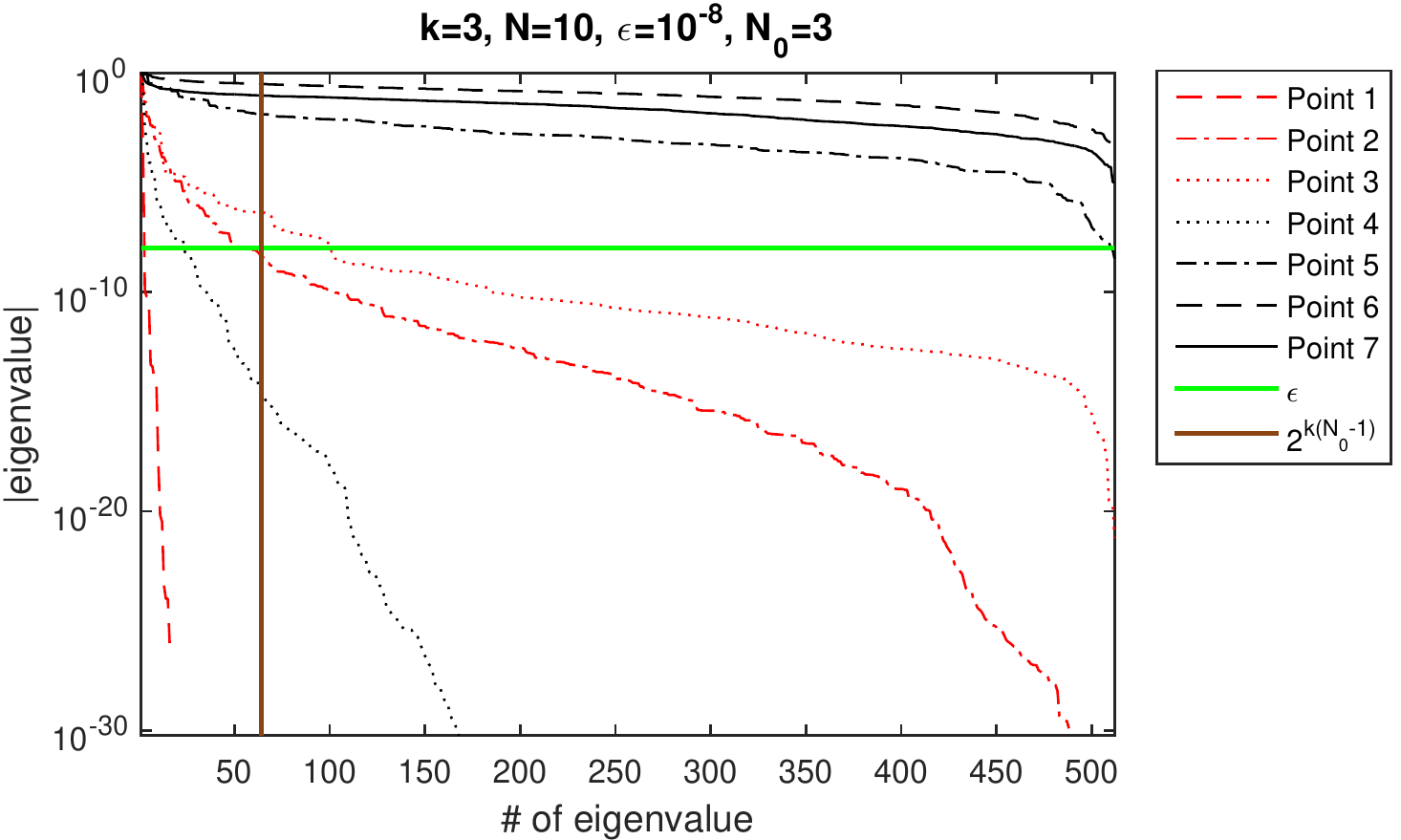}
	\\
	\includegraphics[width=0.49\linewidth]{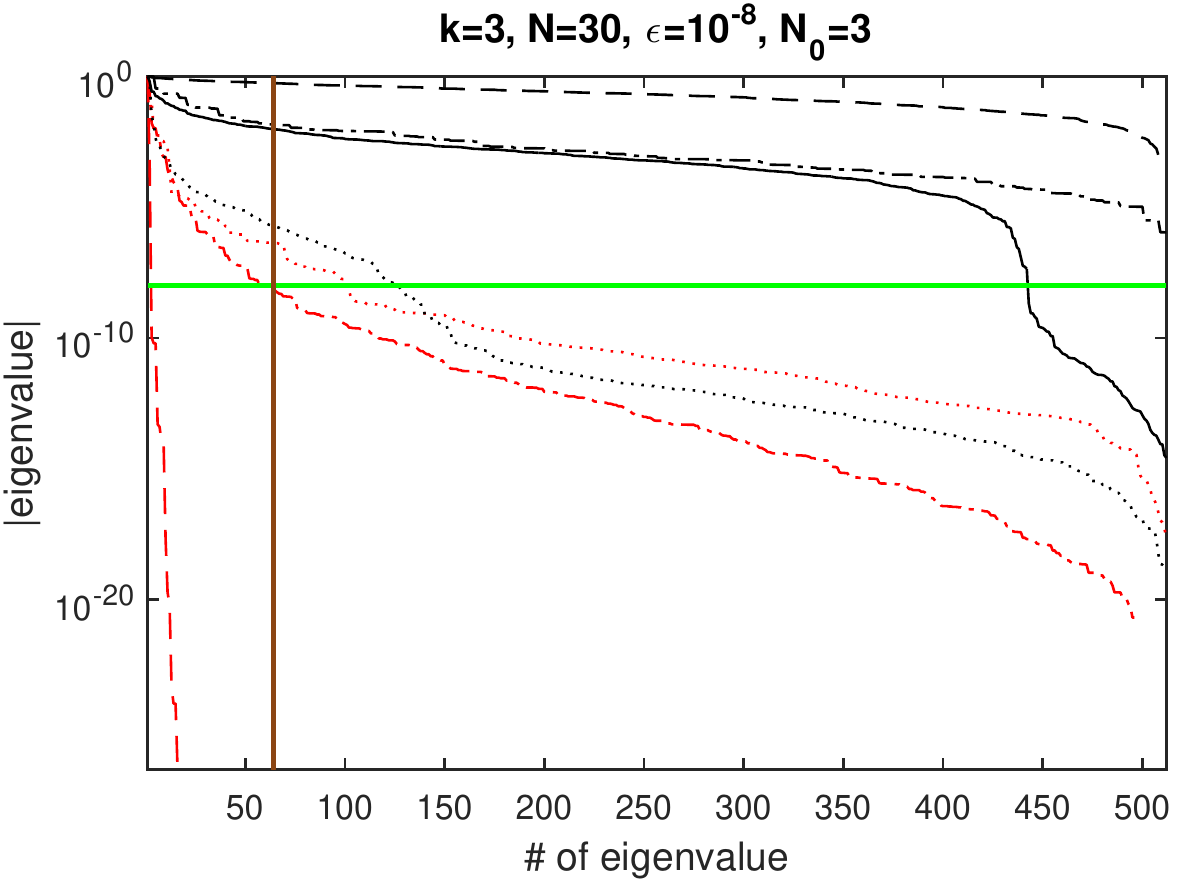}
	\includegraphics[width=0.49\linewidth]{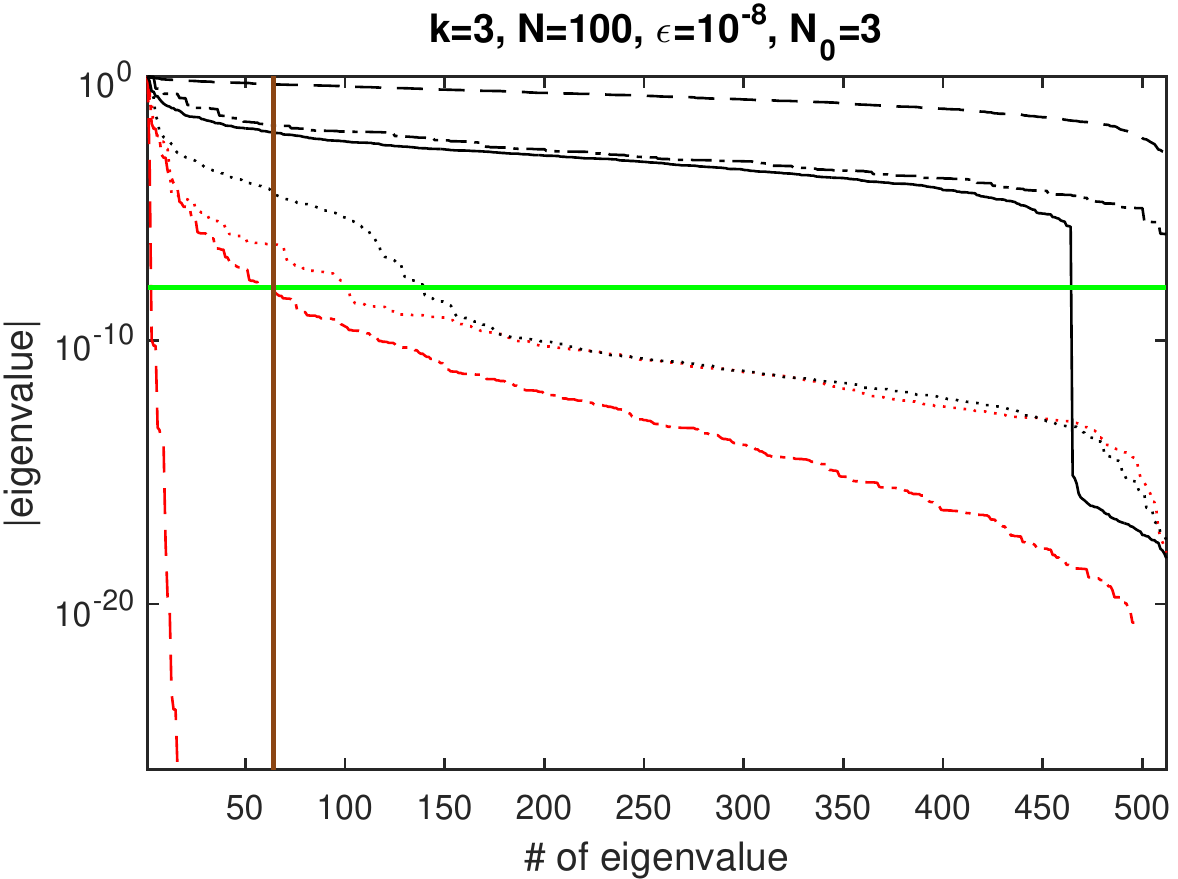}
	\\
	\includegraphics[width=0.49\linewidth]{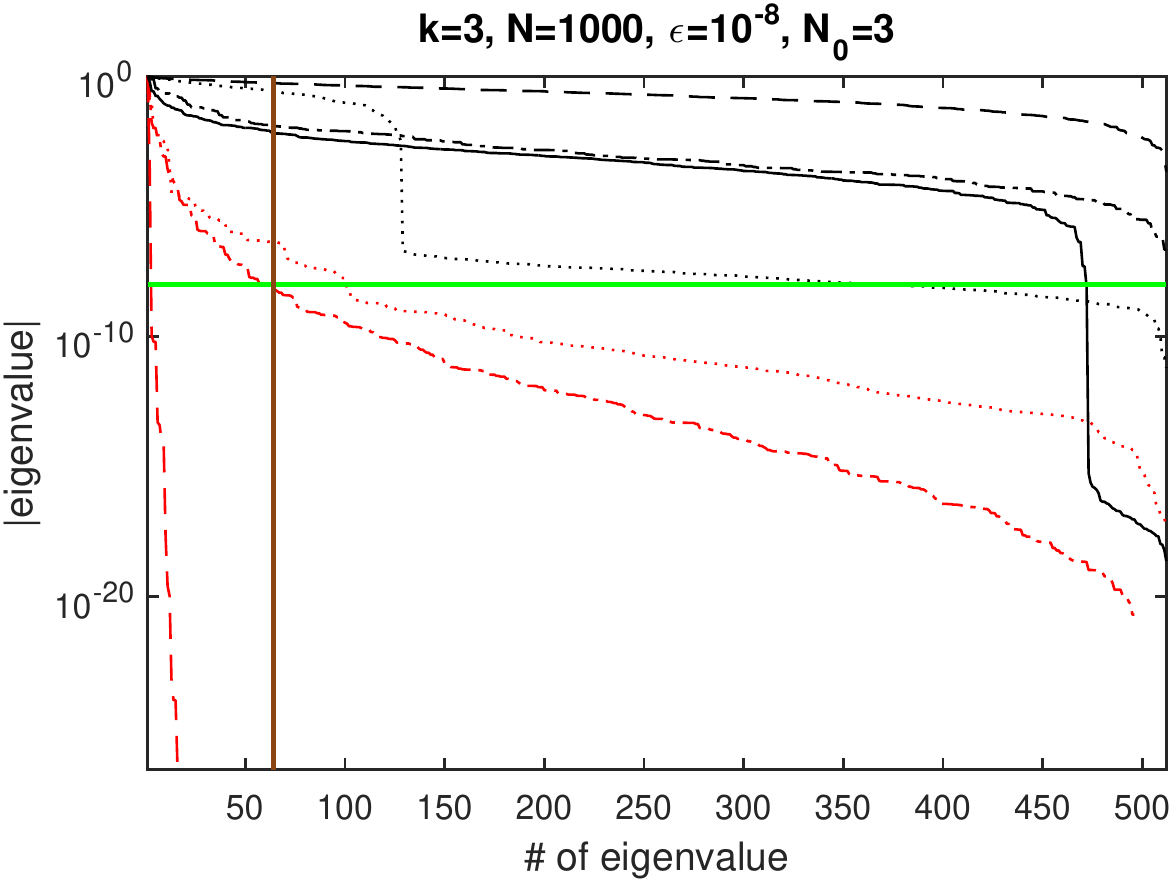}
	\includegraphics[width=0.49\linewidth]{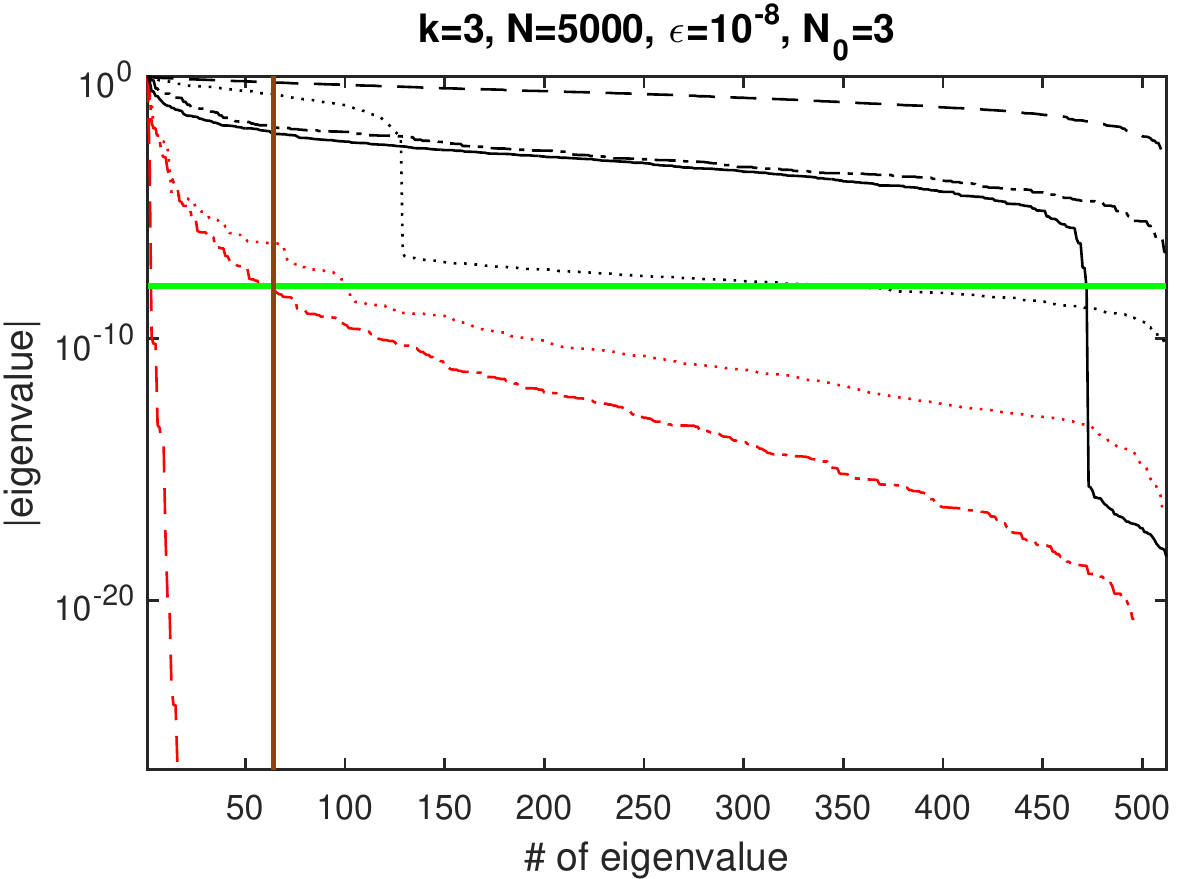}
	\caption[Saturation of the spectra.]{Spectra of $A^{(k)d}_N$ for $k=3$ at $N=10,30,100,1000,5000$ for the same set of points as in figure \ref{fig:Spectra}. We see, that spectra saturate very quickly (before $N=100$). The exception is point 4 where the algorithm encounters numerical instabilities at $N\sim10$ and $N\sim100$ visible in the bottom right diagrams in figure \ref{fig:AsymptBehav}. The saturation of the spectra could possibly be used to speed the algorithm - we could evaluate the recursion formula \eqref{eq:RecursionDiag} only every few steps of iteration and use the same $\kappa$ and $A^{(k)d}_N$ in between. But we have not done that here for precision.}
	\label{fig:SaturationSpectra}
\end{figure}

\begin{figure}[H]
	\centering
	\hspace{1cm}
	\includegraphics[width=0.45\linewidth]{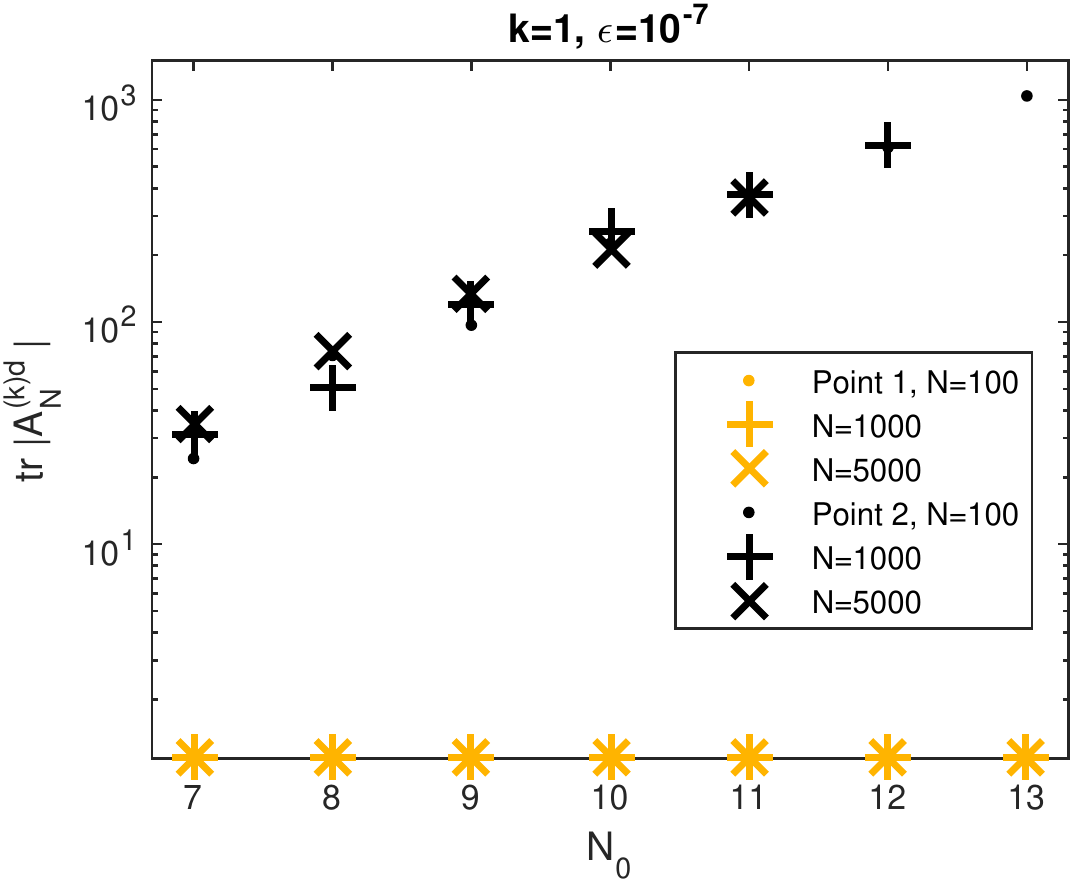}
	\includegraphics[width=0.45\linewidth]{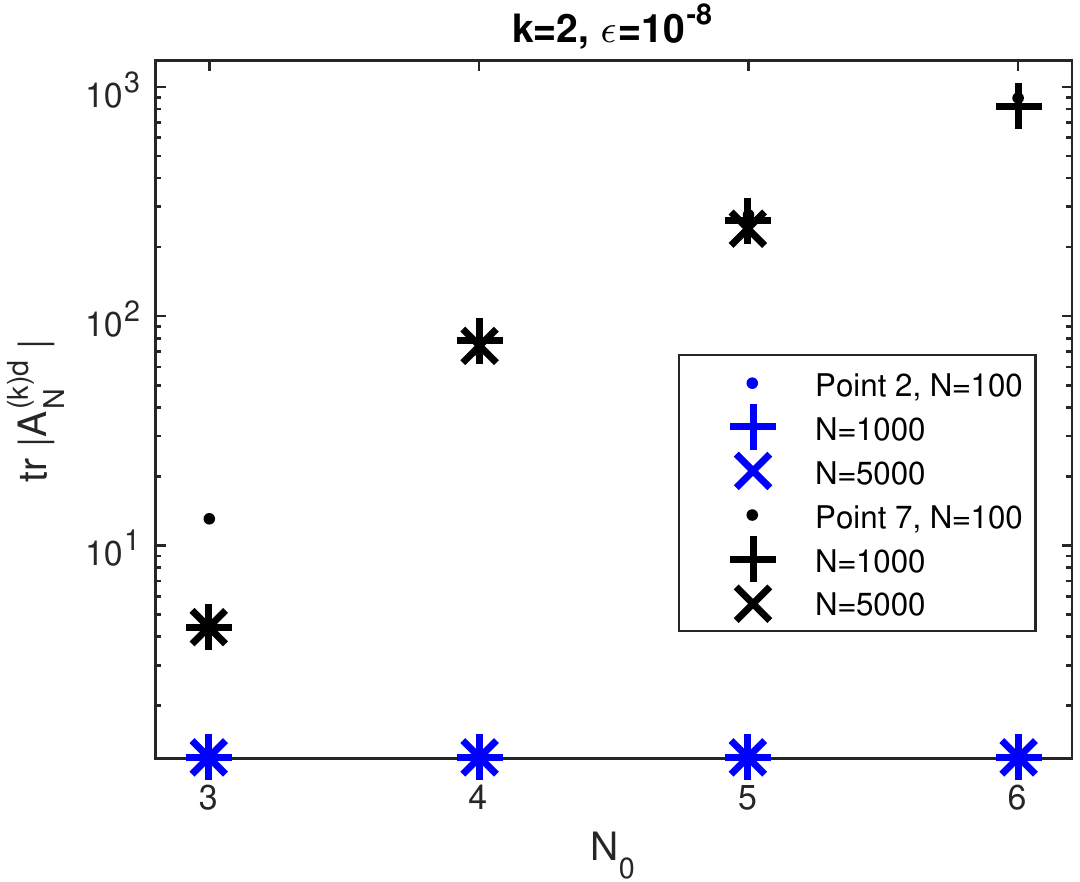}
	\caption[Trace of $\left|A^{(k)d}_N\right|$.]{Semilog plot of spectral magnitudes $\text{tr}\left|A^{(k)d}_N\right|$ in dependence of $N_0$ for $k=1,2$ at $N=100,1000,5000$ for a
	typical ``working" (in colors) and ``non-working" point (black) as labeled in figure \ref{fig:Alpha(h,J)}. We see a clear distinction between the ``working" and ``non-working" phase. Traces of the spectra at the ``working" points are independent of $N_0$ as the truncation is determined by $\epsilon$ there. Traces of the spectra at the ``non-working" points grow exponentially with $N_0$. The exponential growth of the relevant information content means that the algorithm is fundamentally non-renormalizable there.}
	\label{fig:TraceSpectra}
\end{figure}

\subsection{Characteristics of the method}

The reason why the error grows quickly in certain areas of the parameter space and the method does not work there can be understood by looking at the spectra of CTMs, or, diagonal 
matrices $A^{(k)d}_N$. Figure \ref{fig:Spectra} shows the spectra for a chosen set of typical points for $k=2,3$. We see that the spectra visibly divide into two groups according to their steepness - steep spectra corresponding to the points where the algorithm works and flat spectra corresponding to points where the algorithm fails (according to the criteria described in the previous section). Recall that when truncating the corner transfer matrices, we cut away all the eigenvalues smaller than $\epsilon$ but with the additional requirement that no more than $2^{k\left(N_0-1\right)}$ eigenvalues are kept to ensure that the sizes of the matrices in the recursion relation \eqref{eq:RecursionDiag} match. At the ``working"  points, the spectra are steep so the truncation is determined by $\epsilon$. Therefore the cut-away eigenvalues are really small in magnitude and the truncation has a negligible effect on the result. We might say that there the CTMs $\mathcal{A}^{(k)}_{N}$ have effectively a finite rank for any $N$ and we call the method there renormalizable. At the ``non-working" points, however, the spectra are very flat so we are, because of the second criterium, truncating even non-small eigenvalues. As a consequence the error grows rapidly and the method is not reliable. The CTMs have a full rank there and we say that the method is non-renormalizable. We can also see that at some points which are close to the boundary of the ``working" regions of the parameter space, the truncation can be as well determined by the number limit, but the spectra are steep enough so that only small eigenvalues are cut away despite of this.

\begin{figure}[H]
	\includegraphics[width=0.45\linewidth]{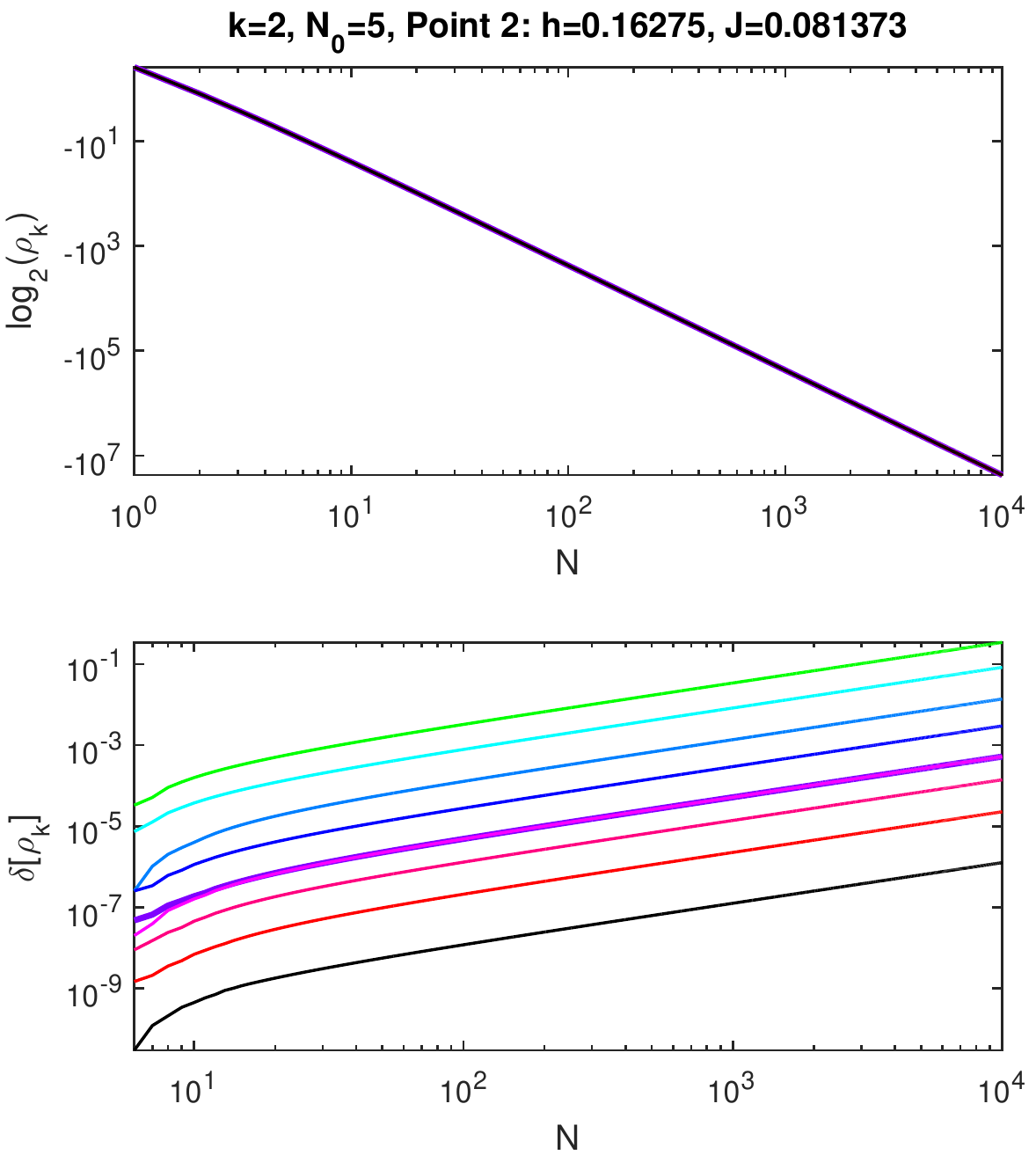}
	\includegraphics[width=0.55\linewidth]{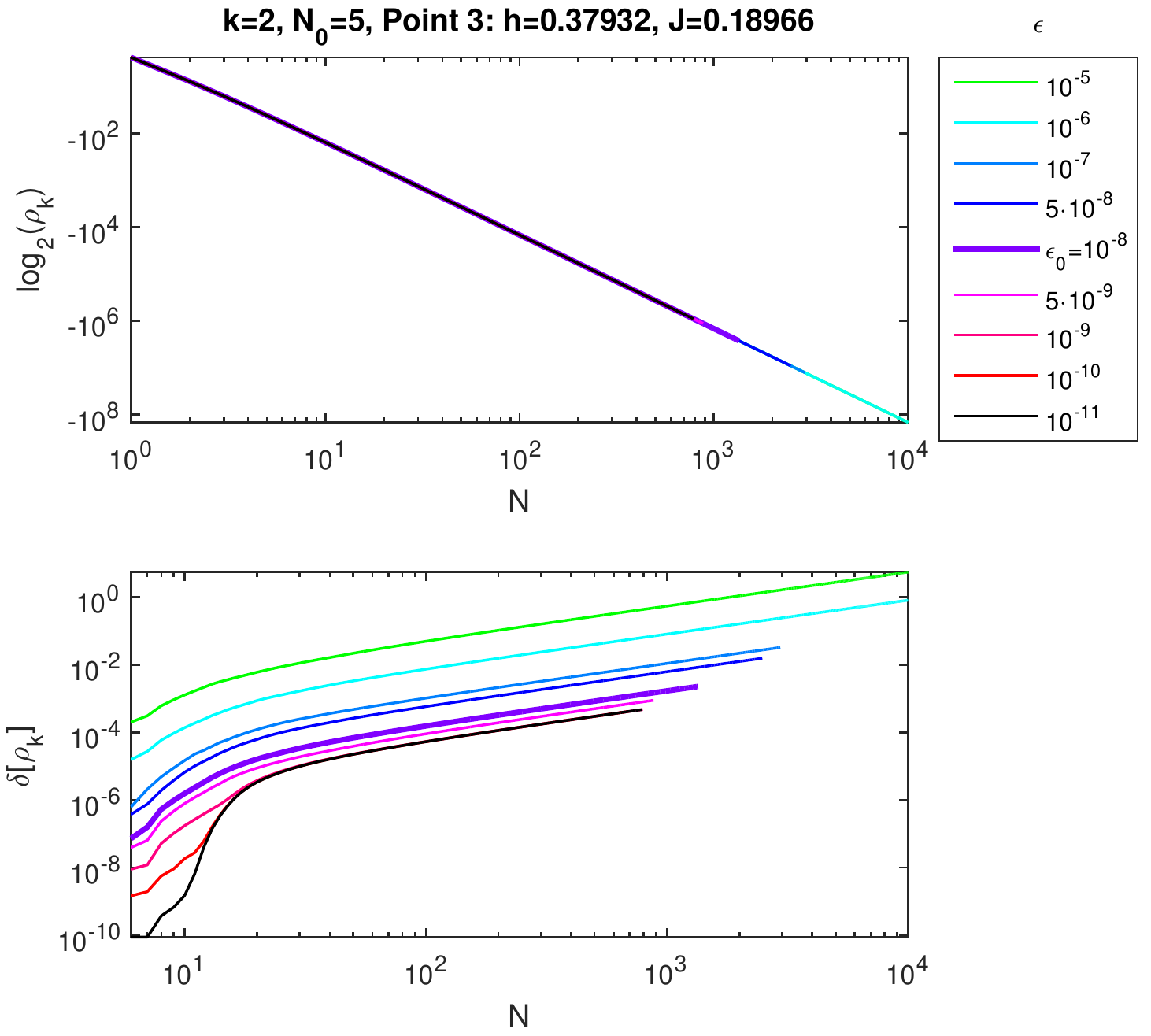} \\
	\includegraphics[width=0.45\linewidth]{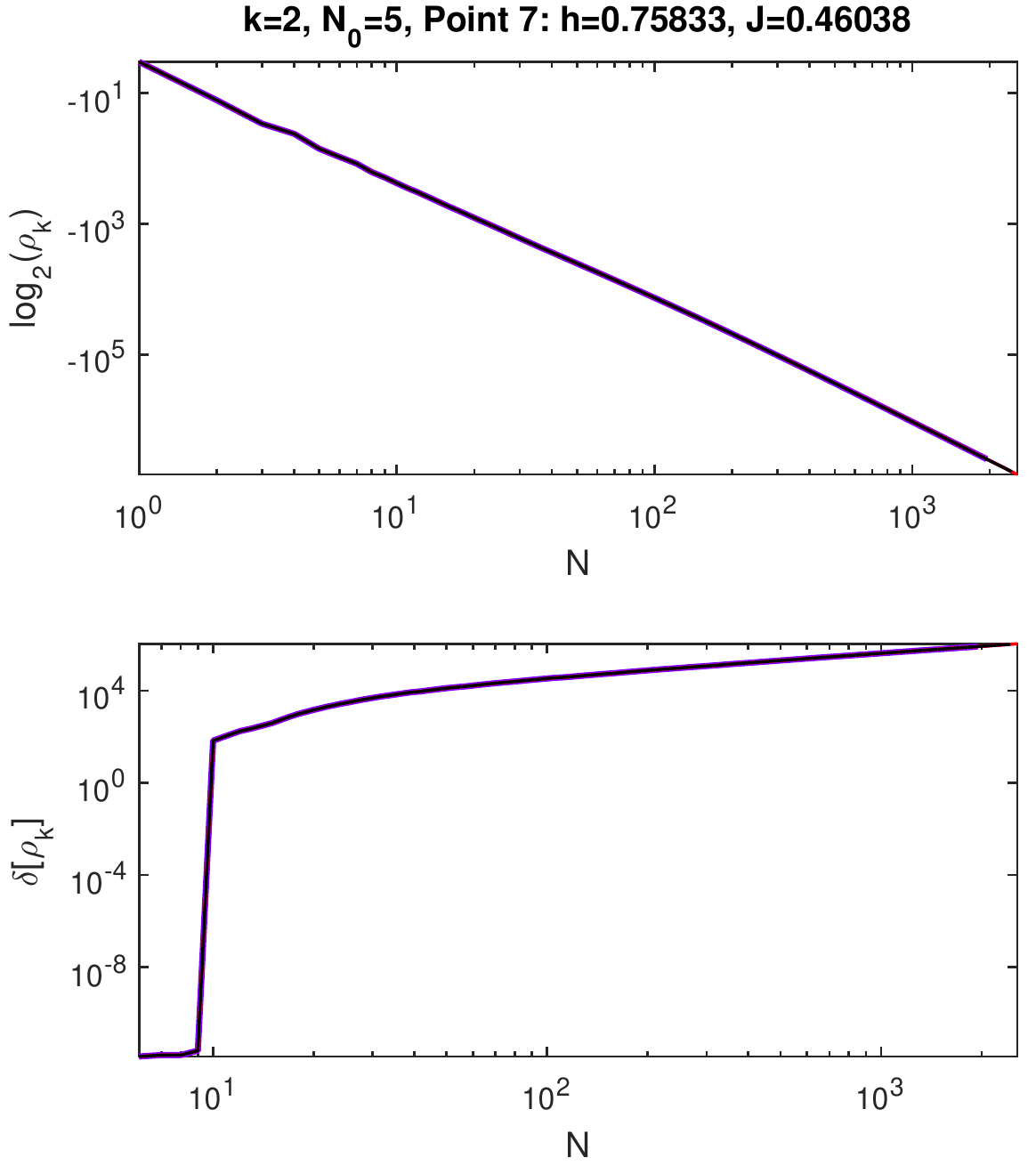}
	\includegraphics[width=0.45\linewidth]{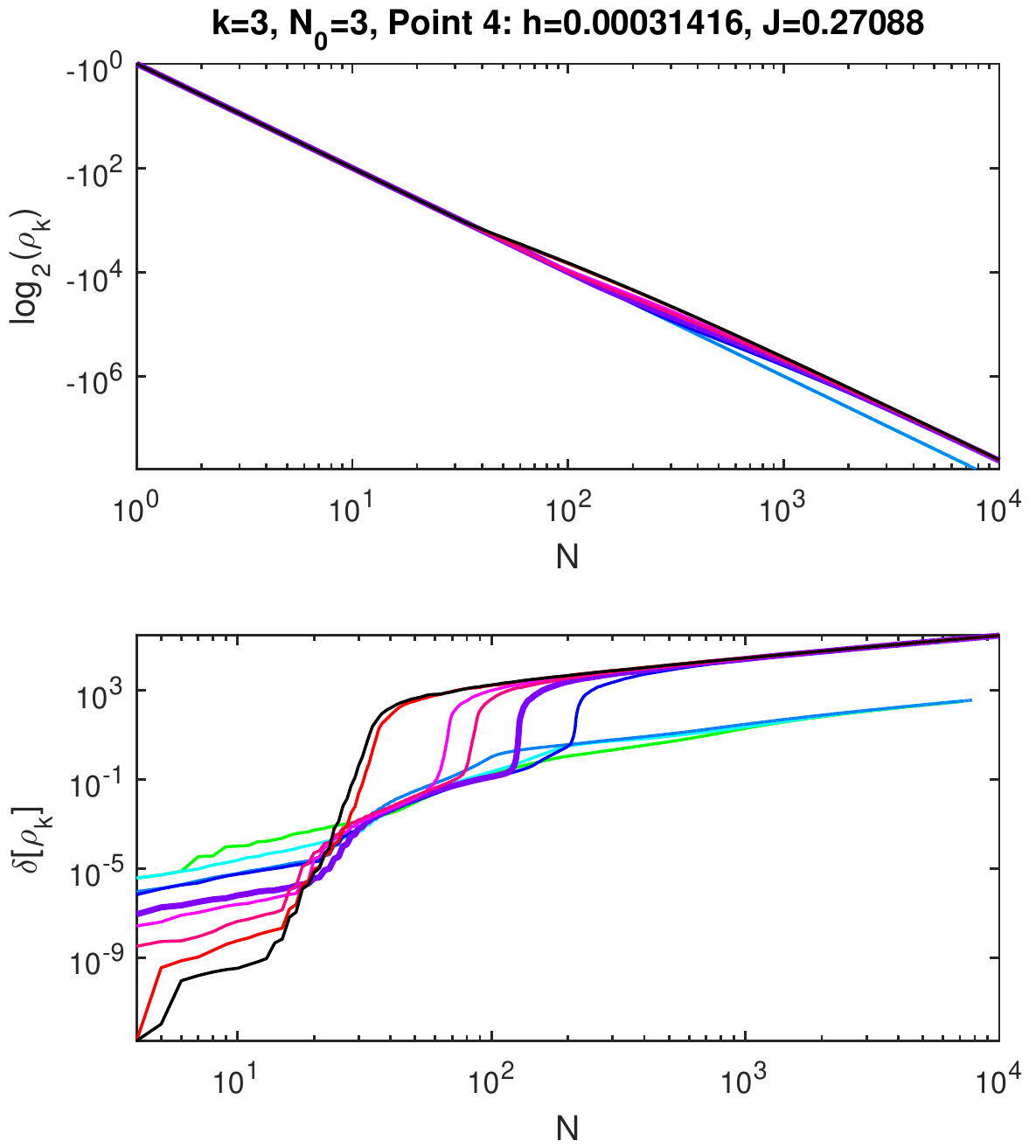}
	\caption[Typical asymptotic behavior.]{Four typical scenarios for scaling of Floquet spectral moments (showing
	$\log_2\left(\rho_k\left(N\right)\right)$) and their relative errors $\delta\left[\rho_k\left(N\right)\right]$. The diagrams are plotted for four among the points denoted in figure \ref{fig:Alpha(h,J)} and the curves are compared for $\epsilon$ ranging over a couple of decades. 
	At the points lying well inside the ``working" areas of the parameter space, as for example in the top left diagrams, the error grows gradually with a rate dependent on the value of $\epsilon$ (the errors only begin to be non-zero at $N=N_0+1$, because the algorithm is exact before). The curves $\rho_k\left(N\right)$ are smooth and independent of $\epsilon$, perfectly following the model \eqref{eq:Model}. At the points close to the boundary of the ``working" regions (top right diagrams), the error grows slightly more rapidly, but remains small and the curves $\rho_k\left(N\right)$ remain well-behaved. At typical points inside the ``non-working" region (bottom left diagrams), the error grows rapidly and this results in the sudden bumps in the curves $\rho_k\left(N\right)$. 
	Also, the curves are independent of $\epsilon$ there because truncation is determined by the eigenvalue number limit. At some points inside the ``non-working" regions (bottom right diagrams) where the algorithm should be working according to the spectra at smaller $N$ (see figure \ref{fig:SaturationSpectra}), the error grows abruptly for smaller values of $\epsilon$. This could be due to numerical instability of division by very small numbers when calculating the inverse of $A_{N-2}^d$ in recursion relation \eqref{eq:RecursionDiag}.
	}
	\label{fig:AsymptBehav}
\end{figure}

\begin{figure}[H]
	\centering
	\includegraphics[width=0.8\linewidth]{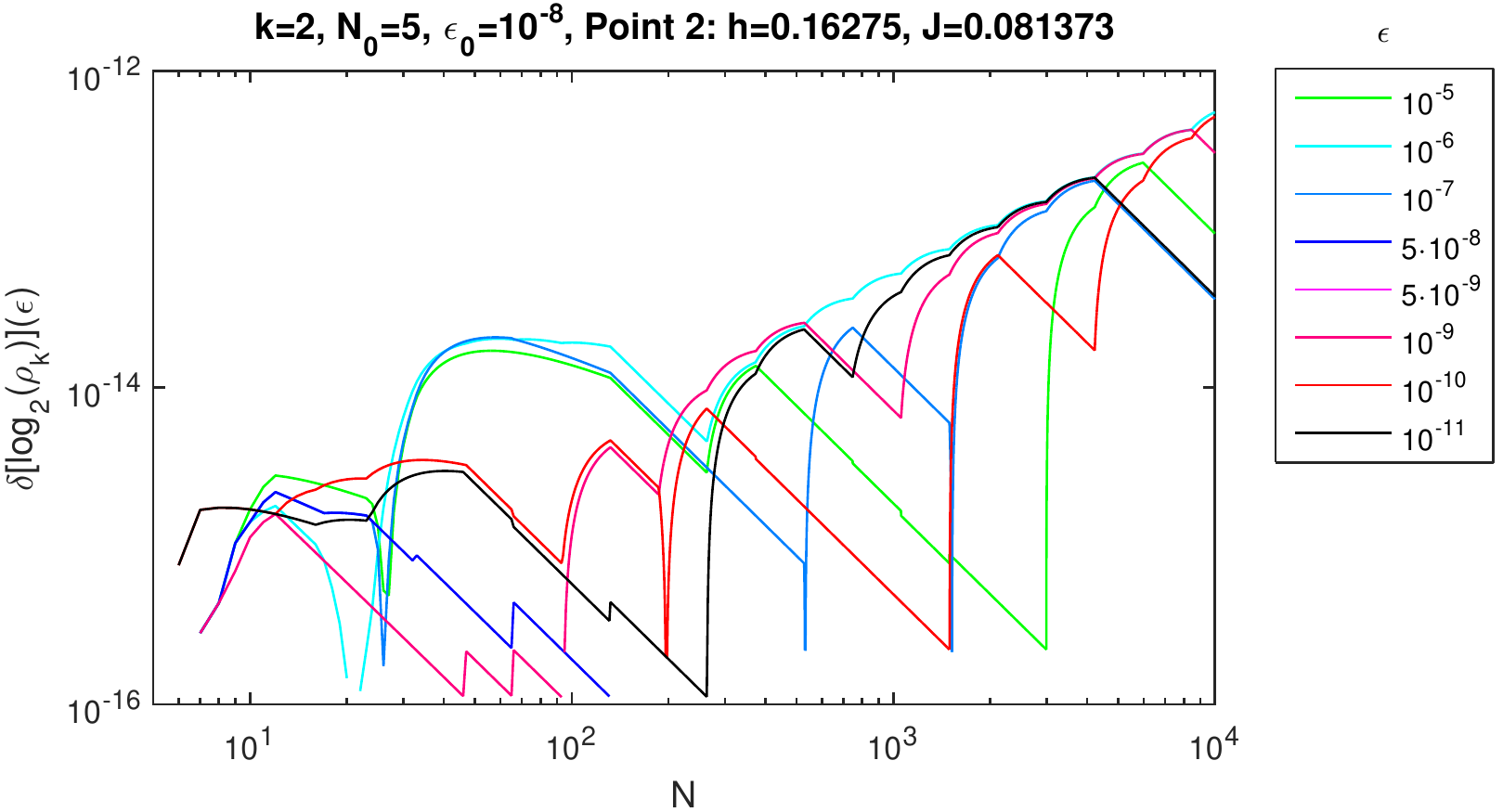}
	\caption[Robustness to $\epsilon$.]{Relative deviation (defined in equation \eqref{eq:RelDeviat}) of the curves $\rho_k\left(N\right)$ computed at the different values of $\epsilon$ from the curves computed at the value $\epsilon_0$ as used for the final results in figure \ref{fig:Alpha(h,J)} for a typical ``working" point. We see that relative deviation is very small. For many ``working" points it is even exactly zero. The small deviations in $\rho_k\left(N\right)$ result in negligible relative deviations of the fitted parameter $\mu_k$ (of the order $\sim10^{-8}-10^{-10}$) so that the method really is very robust to the choice of $\epsilon$.}
	\label{fig:RobustEps}
\end{figure}

Flat spectra in the central region of the parameter space seem to be an insurmountable obstacle preventing the method from working in the full parameter space. The diagrams in the left side of figure \ref{fig:Spectra} give the impression that the ``non-working" regions could be eliminated by increasing $N_0$. However, as shown in the right side of figure \ref{fig:Spectra}, at a larger $N_0$ the spectra at the ``non-working" points just rescale preserving the self-similarity. These points remain ``non-working". The spectra at the ``working" points are unchanged because there, the truncation is determined by $\epsilon$ and so the spectra are independent of $N_0$. Increasing $N_0$ mostly affects the points close to the boundary of the ``working" regions so that their spectra fall within the ``working zone" more reliably. Increasing $N_0$ may therefore make the ``working" regions slightly larger but at the cost of exponentially increasing the computation time and the memory consumption. The ``non-working" regions can probably never be eliminated completely. It would be interesting, however, to understand the physical origin of the flat spectra in the central region of the parameters space. We anticipate that, similarly as in the case of Density Matrix Renormalization Group (DMRG) method \cite{DMRG} (see also \cite{Eisert2010}), the scaling of the entanglement of the auxiliary states given with the CTMs follows an area lay in the ``working" regions a volume law in the ``non-working" regions.

In general, the spectra $A^{(k)d}_N$ saturate very quickly with iteration $N$ (typically before $N=100$) as demonstrated in figure \ref{fig:SaturationSpectra}.

To prove the fundamental difference between the ``working" and the ``non-working" phase, we plot in figure \ref{fig:TraceSpectra} the so-called {\em spectral magnitudes}: total sums of magnitudes of eigenvalues of $A^{(k)d}_N$ in dependence of $N_0$ for a ``working" and ``non-working" point for $k=1,2$. Such spectral magnitudes at the ``working" points are independent of $N_0$ as the truncation is determined by $\epsilon$ there. Spectral magnitudes at the ``non-working" points, however, grow exponentially with $N_0$. The exponential growth of the relevant information content means that the algorithm is fundamentally non-renormalizable there.

The differences in the spectra of $A^{(k)d}_N$ result in different behavior of the curves for scaling of the Floquet spectral moments 
$\rho_k\left(N\right)$ and their relative errors $\delta\left[\rho_k\left(N\right)\right]$. Typical scenarios are presented in figure \ref{fig:AsymptBehav}.

It remains to be shown that the results obtained using the algorithm are robust to the choice of $\epsilon$ at the points where the method is considered to be working. In order to demonstrate that, 
we compute the relative deviation of the curves $\rho_k\left(N\right)$ at different values of $\epsilon$ from the curve obtained with the value $\epsilon_0$ as used for the final results in figure \ref{fig:Alpha(h,J)} (recall that in practice, we are using logarithmized equations (\ref{eq:kappa},\ref{eq:Contraction}) to avoid going below the smallest supported number in the programming language):
\begin{equation}
	\delta\left[\log_2\left(\rho_k\right)\right]\left(\epsilon,N\right):=\left|\frac{\log_2\left(\rho_k\left(\epsilon,N\right)\right)-\log_2\left(\rho_k\left(\epsilon_0,N\right)\right)}{\log_2\left(\rho_k\left(\epsilon_0,N\right)\right)}\right|.\label{eq:RelDeviat}
\end{equation}
For many of the points in the ``working" regions of the parameter space the relative deviation is exactly zero so the results are completely independent of $\epsilon$. When the relative deviation is not exactly zero, it is very small, as presented in figure \ref{fig:RobustEps} for values of $\epsilon$ ranging over a couple of decades. The resulting relative deviations in the fitted parameter $\mu_k$ are of the order $\sim10^{-8}-10^{-10}$ so that the method really is robust to the choice of $\epsilon$.

\section{Conclusions}
\label{sec:Conc}
We have developed a renormalizable, numerically exact method for the computation of quasienergy spectra of 2D strongly-coupled many-body Floquet systems based on Baxter's corner transfer matrices. We have demonstrated its functionality exemplified by the kicked 2D quantum Ising model. The method enabled us to compute the first few Fourier coefficients, $\rho_k$, of the quasienergy spectrum for the systems of arbitrary size. We performed the computation for up to $10\,000\times10\,000$ spin lattices which was more than enough to determine the system's properties in the the thermodynamic limit. The phase diagrams $\rho_k\left(h,J\right)$ for small (finite) lattices agreed with the diagrams obtained in Ref.~\cite{PinedaProsen2014}. The decay of $\rho_k(N)$ qualitatively agreed in the leading order with the theoretical prediction obtained using random matrix theory, consistent with the hypothesis that the system is quantum chaotic. The decay rate, however, was highly and nontrivially dependent on the system's parameters $h$,$J$. We presented rich phase diagrams of the leading coefficient. We have practically demonstrated the robustness of the method to the choice of the truncation parameter $\epsilon$ in the regions of the parameter space where the method is renormalizable. What is perhaps the most interesting discovery is that the method is renormalizable only in certain regions of the parameter space and nonrenormalizable in the central region due to flat spectra of the corner transfer matrices. It would be interesting to find the physical origin for this phenomenon. Based on the experience with Density Matrix Renormalizaiton Group, we expect the steepness of the spectra, or the effective rank of corner transfer matrices, to be related to quantum entanglement and correlations. Our results might point to an interesting new type of quantum phase transition from area- to volume-like entanglement entropy scaling in 2D quantum lattice systems. We have also computed the phase diagrams at different angles of the external magnetic field $\theta$ and found a rich and interesting behavior but do not discuss it here for brevity.

The CTM-based numerical method should be in principle widely applicable for studying periodically driven strongly coupled 2D quantum many-body systems. In the future it would be interesting to find more applications, explore whether the corner transfer matrices are renormalizable for a general system and find the connection between renormalizability and entanglement entropy. 

\section*{Acknowledgements}

We thank Bor Plestenjak and Boris Gutkin for useful discussions, Nastja Miheljak for graphic design and acknowledge financial support by Slovenian Research Agency (ARRS) under grants P1-0044, J1-5439 and N1-0025.

\section*{References}
\bibliographystyle{unsrt}
\bibliography{CTMpaper}

\begin{thebibliography}{10}

\bibitem{Eisert2010}
J.~Eisert, M.~Cramer, and M.~B. Plenio.
\newblock \textit{Colloquium} : Area laws for the entanglement entropy.
\newblock {\em Rev. Mod. Phys.}, 82:277--306, Feb 2010.

\bibitem{Haake}
F.~Haake.
\newblock {\em Quantum {S}ignatures of {C}haos}.
\newblock Springer-{V}erlag, {T}hird edition, 2010.

\bibitem{ProsenPRL1998}
T.~Prosen.
\newblock Time evolution of a quantum many-body system: Transition from
  integrability to ergodicity in the thermodynamic limit.
\newblock {\em Phys. Rev. Lett.}, 80:1808--1811, Mar 1998.

\bibitem{ProsenJPA1998}
T.~Prosen.
\newblock Quantum invariants of motion in a generic many-body system.
\newblock {\em Journal of Physics A: Mathematical and General}, 31(37):L645,
  1998.

\bibitem{ProsenPRE1999}
T.~Prosen.
\newblock Ergodic properties of a generic nonintegrable quantum many-body
  system in the thermodynamic limit.
\newblock {\em Phys. Rev. E}, 60:3949--3968, Oct 1999.

\bibitem{ProsenPRE2002}
T.~Prosen.
\newblock General relation between quantum ergodicity and fidelity of quantum
  dynamics.
\newblock {\em Phys. Rev. E}, 65:036208, Feb 2002.

\bibitem{Polkovnikov}
M.~Bukov, L.~D'Alessio, and A.~Polkovnikov.
\newblock Universal high-frequency behavior of periodically driven systems:
  from dynamical stabilization to floquet engineering.
\newblock {\em Advances in Physics}, 64(2):139--226, 2015.

\bibitem{Rigol}
L.~D'Alessio and M.~Rigol.
\newblock Long-time behavior of isolated periodically driven interacting
  lattice systems.
\newblock {\em Phys. Rev. X}, 4:041048, Dec 2014.

\bibitem{Demler}
R.~Citro, E.~G.~Dalla Torre, L.~D’Alessio, A.~Polkovnikov, M.~Babadi, T.~Oka,
  and E.~Demler.
\newblock Dynamical stability of a many-body kapitza pendulum.
\newblock {\em Annals of Physics}, 360:694 -- 710, 2015.

\bibitem{Kuwahara2015}
T.~Kuwahara, T.~Mori, and K.~Saito.
\newblock Floquet-magnus {T}heory and {G}eneric {T}ransient {D}ynamics in
  {P}eriodically {D}riven {M}any-body {Q}uantum {S}ystems.
\newblock {\em arXiv:1508.05797}, 2015.

\bibitem{Mori2015}
T.~Mori, T.~Kuwahara, and K.~Saito.
\newblock Rigorous bound on energy absorption and generic relaxation in
  periodically driven quantum systems.
\newblock {\em arXiv:1509.03968}, 2015.

\bibitem{Abanin2015}
D.~A. Abanin, W.~De Roeck, and W.~W. Ho.
\newblock Effective {H}amiltonians, prethermalization and slow energy
  absorption in periodically driven many-body systems.
\newblock {\em arXiv:1510.03405}, 2015.

\bibitem{PinedaProsen2014}
C.~Pineda, T.~Prosen, and E.~Villase{\~n}or.
\newblock Two dimensional kicked quantum {I}sing model: dynamical phase
  transitions.
\newblock {\em New Journal of Physics}, 16:123044, 2014.

\bibitem{MehtaRM}
M.~L. Mehta.
\newblock {\em Random {M}atrices}.
\newblock Elsevier, {T}hird edition, 2004.

\bibitem{BaxterESM}
R.~J. Baxter.
\newblock {\em Exactly {S}olved {M}odels in {S}tatistical {M}echanics}.
\newblock Academic {P}ress, 1989.

\bibitem{BaxterCTM1981}
R.~J. Baxter.
\newblock Corner {T}ransfer {M}atrices.
\newblock {\em Physica A}, 106:18--27, 1981.

\bibitem{Orus2012}
R.~Orús.
\newblock Exploring corner transfer matrices and corner tensors for the
  classical simulation of quantum lattice systems.
\newblock {\em Physical Review B}, 85:205117, 2012.

\bibitem{Bartel2008}
E.~Bartel and A.~Schadschneider.
\newblock Quantum {C}orner — {T}ransfer {M}atrix {DMRG}.
\newblock {\em International Journal of Modern Physics C}, 19(08):1145--1161,
  2008.

\bibitem{DMRG}
U.~Schollwock.
\newblock The density-matrix renormalization group in the age of matrix product
  states.
\newblock {\em Annals of Physics}, 326(1):96 -- 192, 2011.
\newblock January 2011 Special Issue.

\end{thebibliography}

\end{document}